\documentclass[12pt]{article}
\usepackage[toc,page]{appendix}
\usepackage{xcolor}
\usepackage{tikz}
\usetikzlibrary{cd}
\usepackage[compat=1.1.0]{tikz-feynman} \tikzfeynmanset{warn luatex=false}
\usepackage{mathtools}
\usepackage{amsthm,amsmath,latexsym,amssymb,amsfonts,amscd}
\usepackage{graphics,lscape,fancyhdr,array,stmaryrd,euscript,wrapfig}
\usepackage{empheq,slashed}
\usepackage{verbatim,slashed}
\numberwithin{equation}{section}
\usepackage{hyperref,setspace}
\usepackage{dsfont}
\usepackage{mathrsfs}

\usepackage[numbers,sort&compress]{natbib}
\usepackage[nottoc]{tocbibind}

\newcommand{\be}{\begin{equation}}
\newcommand{\ee}{\end{equation}}

\newcommand{\fdu}[2]{{}_{#1}{}^{#2}\,}

\newcommand{\besubeqs}{\begin{subequations}}
\newcommand{\esubeqs}{\end{subequations}}

\newcommand{\PP}{{\mathbb{P}}}

\newcommand{\PPb}{{\bar{\mathbb{P}}}}

\newcommand{\fA}{\mathfrak{f}}   
\newcommand{\fB}{\mathbf{f}}     

\newcommand{\definition}{\mathrel{\mathop:}=}

\begin{document}
\pagenumbering{gobble}
\hfill
\vskip 0.01\textheight
\begin{center}
{\Large\bfseries Associativity of celestial OPE, higher spins\\
[5mm]
and self-duality}

\vspace{0.4cm}

\vskip 0.03\textheight
\renewcommand{\thefootnote}{\fnsymbol{footnote}} Mattia Serrani${}^{\pi}$
\renewcommand{\thefootnote}{\arabic{footnote}}
\vskip 0.03\textheight
\centering
\href{mailto:mattia.serrani@umons.ac.be}{\texttt{mattia.serrani@umons.ac.be}}
\vskip 0.03\textheight

{\em ${}^{\pi}$ Service de Physique de l'Univers, Champs et Gravitation, \\ Universit\'e de Mons, 20 place du Parc, 7000 Mons, 
Belgium}
\end{center}

\vskip 0.02\textheight

\begin{abstract}
We highlight and clarify the connection between several ideas and self-dual theories: (a) the operator product expansion (OPE) associativity in celestial conformal field theory (CCFT); (b) the vanishing of tree-level amplitudes; (c) the Jacobi identity for the ``gauge'' algebra; (d) the light-cone holomorphic constraints. Naturally, (b), (c), or (d) are closely related to self-duality. In particular, the recently classified \cite{Serrani:2025owx} chiral higher-spin theories with one- and two-derivative interactions (i.e. with gauge and gravitational interactions, which are extensions of self-dual Yang-Mills and self-dual gravity) also satisfy the OPE associativity constraint. We discuss the OPE associativity constraint and the holomorphic constraint for the most general class of cubic vertices.

\end{abstract}

\newpage
\tableofcontents
\newpage

\section{Introduction}\label{Paper2_section1}
\pagenumbering{arabic}
\setcounter{page}{2}

In the paper, we elaborate on the links between several recent and old developments. (1) The requirement of associativity of the celestial OPE \cite{Pate:2019lpp,Himwich:2021dau,Strominger:2021mtt,Costello:2022upu,Bittleston:2022jeq} leads to certain constraints for couplings \cite{Mago:2021wje} that relate low- and higher-derivative ones. Such constraints have nothing to do with Lorentz invariance in the bulk and are puzzling at first sight. (2) It was shown in \cite{Ren:2022sws,Monteiro:2022lwm} that, remarkably, Chiral higher-spin gravity passes the celestial OPE associativity test. (3) Chiral higher-spin gravity is, under certain assumptions, a unique perturbatively local higher-spin gravity in $4d$ \cite{Metsaev:1991mt,Metsaev:1991nb,Ponomarev:2016lrm,Ponomarev:2017nrr}. (4) It was shown in \cite{Ponomarev:2017nrr} that the equations of Chiral higher-spin gravity can be formulated as a self-duality constraint for a certain ``gauge'' algebra. Two new theories were also found in \cite{Ponomarev:2017nrr} that can be thought of as higher-spin extensions of self-dual Yang-Mills (SDYM) and self-dual gravity (SDGR), which are contractions of Chiral higher-spin gravity that feature only gauge and gravitational interactions,\footnote{As will be made clearer later, by gauge and gravitational interactions, we mean those that have one- and two-derivative cubic interactions, respectively, when counted in the light-cone gauge.} respectively. (5) It is a folklore statement that self-dual theories have vanishing tree-level amplitudes, which is one of the definitions of classical integrability. (6) Very recently, it was shown \cite{Serrani:2025owx} that there are more ``chiral'' theories with higher-spin fields, all of which activate different subsets of the Chiral higher-spin gravity couplings. In the paper, we aim to clarify the relation between all these statements, which can be summarised by the following diagram 
$$ 
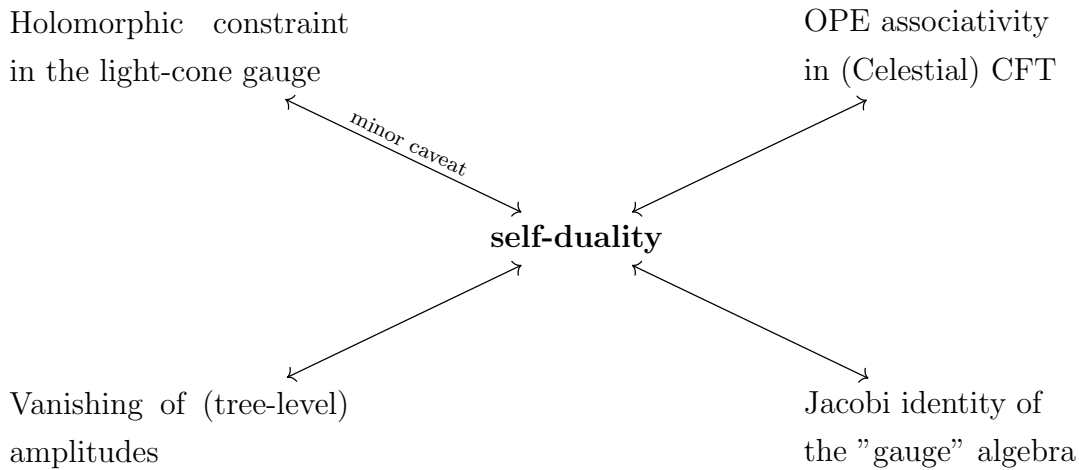
\begin{tikzcd}[row sep=1.5cm, column sep=1.5cm]
\parbox{4.5cm}{Holomorphic constraint in the light-cone gauge} \arrow[dr, leftrightarrow, "\text{minor caveat}", sloped, above] & & \parbox{4.5cm}{OPE associativity \\ in (Celestial) CFT} \arrow[dl, leftrightarrow] \\
& \textbf{self-duality} & \\
\parbox{4.5cm}{Vanishing of (tree-level) amplitudes} \arrow[ur, leftrightarrow] & & \parbox{4.5cm}{Jacobi identity of \\ the ''gauge'' algebra} \arrow[ul, leftrightarrow]
\end{tikzcd} 
$$
In more detail, let us begin with higher-spin gravity \cite{Bekaert:2022poo,Ponomarev:2022vjb}, which was the initial motivation for this work. Chiral higher-spin gravity and its contractions \cite{Metsaev:1991mt,Metsaev:1991nb,Ponomarev:2016lrm,Ponomarev:2017nrr} were first constructed in the light-cone gauge, relying on the earlier works by Bengtsson, Bengtsson, and Brink \cite{Bengtsson:1983pd,Bengtsson:1983pg}, who were the first to construct cubic interactions of massless higher-spin fields. The light-cone gauge has always been a very useful tool for initial developments, with the main idea being to directly construct generators of the Poincaré algebra order by order in terms of the physical degrees of freedom, putting aside any gauge redundancy a covariant formulation can lead to. A remarkable feature \cite{Metsaev:1991mt,Metsaev:1991nb} of the light-cone gauge in $4d$ is that the quartic constraint, which is equivalent to the Poincaré invariance of the quartic amplitude, contains the (anti)holomorphic parts that are insensitive to the quartic Hamiltonian itself, but still heavily constrain the cubic couplings. The (anti)holomorphic constraints involve only the vertices whose cubic amplitudes are made of (square)angle brackets. Solving the holomorphic constraint leads to Chiral higher-spin gravity, whose cubic amplitudes are 
\begin{align}
    \frac{1}{\Gamma[\lambda_1+\lambda_2+\lambda_3]}[12]^{\lambda_1+\lambda_2-\lambda_3}[23]^{\lambda_2+\lambda_3-\lambda_1}[13]^{\lambda_1+\lambda_3-\lambda_2}\,, \qquad \sum\lambda_i>0\,.
\end{align}
The Gamma-function is what the holomorphic constraint fixes the helicity dependence of couplings to be. This is a unique solution provided one starts with at least one nonabelian higher-spin self-interaction, as proven in \cite{Serrani:2025owx}. It should be noted that self-dual Yang-Mills (SDYM) and self-dual gravity (SDGR) are also solutions to the holomorphic constraint. As found in \cite{Ponomarev:2016lrm,Ponomarev:2017nrr,Monteiro:2022xwq,Serrani:2025owx}, there are other solutions to the holomorphic constraint that feature only gauge or gravitational interactions (HS-SDGR and HS-SDYM) and can even have finitely many higher-spin fields. In fact, all solutions of the holomorphic constraint immediately give consistent theories by setting all higher orders and anti-holomorphic couplings to zero. In this sense, Chiral higher-spin gravity is the maximal completion of self-dual theories that activate all spins and all couplings. All of these theories --- with a minor caveat to be discussed --- have vanishing tree-level amplitudes, which can also be understood as being closely related to self-duality. 

The relation of the above to self-duality can be clarified with the help of the kinematic and ``gauge'' (Lie) algebra \cite{Monteiro:2011pc,Ponomarev:2017nrr,Ponomarev:2024jyg}.\footnote{In \cite{Monteiro:2011pc}, the kinematical algebra of SDYM is identified as the Lie algebra defined by the SDYM cubic vertex after stripping off the colour factor. By contrast, in \cite{Ponomarev:2017nrr,Ponomarev:2024jyg}, the “gauge” algebra is defined as the Lie algebra obtained by factoring out the SDYM vertex from the cubic vertices.} It is a nontrivial statement that there is a Lie algebra behind any of the theories above, which amounts to checking the Jacobi identity. The gauge algebra allows one to rewrite the equations of motion as a self-duality constraint. Remarkably, all chiral higher-spin theories (with a minor caveat), including the new ones \cite{Serrani:2025owx}, have a gauge algebra, i.e. are self-dual, as we show in the paper. 

We also show that all chiral higher-spin theories  --- with a minor caveat ---  pass the celestial OPE associativity test with the first example given in \cite{Ren:2022sws,Monteiro:2022lwm}. We also find a general solution to the celestial OPE associativity and provide some explicit low-spin examples. The same condition can also be related to the vanishing of the four-point amplitudes. 

The paper is organised as follows: In Section \ref{Paper2_section2}, we review the light-front approach to higher-spin interactions, and in particular the quartic holomorphic constraint. In Section \ref{Paper2_section3}, we review the associativity of the celestial OPE and extend it to the case where fields live in some representation of a gauge group $G$. In Section \ref{Paper2_section4}, we solve the celestial OPE associativity constraint in various cases by rewriting it in terms of the
light-cone variables. In Section \ref{Paper2_section5}, we discuss in detail lower-spin theories that satisfy the OPE associativity constraint. We also match the results with those in \cite{Ren:2022sws} and identify additional solutions. In Section \ref{Paper2_section6}, we discuss the relations between: (a) the celestial OPE associativity; (b) the vanishing of tree-level amplitudes; (c) the Jacobi identity for the gauge algebra; (d) the light-cone holomorphic constraints. Finally, in Section \ref{Paper2_section7}, we conclude and outline possible future directions for extending the connection between the light-cone constraints and the OPE associativity in CCFT.

We include four appendices. Appendix \ref{Paper2_AppendixA} introduces our light-cone notation. Appendix \ref{Paper2_AppendixB} presents the CCFT notation for massless fields. In Appendix \ref{Paper2_AppendixC}, we present an on-shell map between the $4d$ light-cone and $2d$ CCFT formalisms via their relation to spinor-helicity. In Appendix \ref{Paper2_AppendixD}, we solve the OPE associativity constraint for the most general class of cubic vertices.

\section{Light-cone holomorphic constraint}
\label{Paper2_section2}

We begin by briefly reviewing the essentials of the light-front approach to interactions and the derivation of the \textit{quartic holomorphic constraint}. For further details and background, see \cite{Ponomarev:2016lrm,Ponomarev:2017nrr,Ponomarev:2016cwi,Serrani:2025owx,Ponomarev:2022vjb}. The conventions for the light-cone notation adopted are detailed in Appendix~\ref{Paper2_AppendixA}.

The light-front deformation procedure relies on two key ingredients: the Hamiltonian formulation of classical/quantum field theory in light-cone gauge and the non-linear realisation of the Poincaré algebra through the inclusion of higher-order corrections to the free (quadratic) generators of the Poincaré algebra. Accordingly, we begin with the $4d$ Poincaré algebra $iso(3,1)$, defined as
\begin{subequations}\label{Poincare1}
\begin{align}
    [P^A,P^B]=&\,0\,,\\
    [J^{AB},P^C]=&\,P^A\eta^{BC}-P^B\eta^{AC}\,,\\
    [J^{AB},J^{CD}]=&\,J^{AD}\eta^{BC}-J^{BD}\eta^{AC}-J^{AC}\eta^{BD}+J^{BC}\eta^{AD}\,,
\end{align}
\end{subequations}
where $P^A$ are the generators of translations and $J^{AB}$ of Lorentz transformations.

In $4d$, massless spinning fields have two degrees of freedom; i.e. they are effectively represented by two ``scalars'' --- except for the scalar field. We denote $\phi^{\lambda}$ and $\phi^{-\lambda}$ as the helicity $+\lambda$ and $-\lambda$ fields, respectively. In Lorentzian signature, the one adopted here, they are complex fields and complex conjugates of each other $\phi^{-\lambda}=(\phi^{\lambda})^*$. The free action is $S=\tfrac{1}{2}\int d^4x\, \phi^{-\lambda}\Box\phi^{\lambda}$ and is real in any signature. 

We work with Fourier transformed fields with respect to $x^-$ and the transverse directions $x$ and $\bar{x}$. The Dirac bracket is given by
\begin{equation}
    [\phi_q^{\lambda}(x^+),\phi_p^{s}(x^+)]=\delta^{\lambda,-s}\frac{\delta^3(q+p)}{2q^+}\,.
\end{equation}
In the Hamiltonian approach, it is crucial to distinguish between the kinematical and dynamical generators of the Poincaré algebra \eqref{Poincare1}. The kinematical generators correspond to the stability group of the codimension-one hypersurface on which quantisation is performed (i.e. in the light-front at $x^+=0$), and they remain unaffected by the introduction of interactions. In contrast, the dynamical generators are deformed once interactions are included, as they govern the evolution from one hypersurface to another. The allowed deformations are determined by solving the dynamical constraints order by order in the deformation procedure while carefully imposing locality at each order.

The free field realisation of the kinematical Poincaré generators reads\footnote{Following standard notations in the light-cone, we rename $\beta=p^+$.}
\begin{subequations}
\begin{align}
    &P^+=\beta\,,&
    &P=q\,,&
    &\bar{P}=\bar{q}\,,\\
    &J^{x+}=-\beta\frac{\partial}{\partial \bar{q}}\,,&
    &J^{\bar{x}+}=-\beta\frac{\partial}{\partial q}\,,&
    &J^{-+}=-N_{\beta}-1\,,\\
    &J^{x\bar{x}}=N_q-N_{\bar{q}}-\lambda\,,
\end{align}
\end{subequations}
where $N_q=q\partial_q$ is the Euler operator.
The (free) dynamical generators are
\begin{align}
    &H_2=-\frac{q\bar{q}}{\beta}\,,&
    &J_2^{z-}=\frac{\partial}{\partial\bar{q}}\frac{q\bar{q}}{\beta}+q\frac{\partial}{\partial\beta}+\lambda\frac{q}{\beta}\,,&
    &J_2^{\bar{z}-}=\frac{\partial}{\partial q}\frac{q\bar{q}}{\beta}+\bar{q}\frac{\partial}{\partial\beta}-\lambda\frac{\bar{q}}{\beta}\,.
\end{align}
We now deform the dynamical generators using a local ansatz:\footnote{Only the dynamical generators are deformed. One advantage of working in light-front quantisation is that the number of dynamical generators attains its minimum. Out of $10$ Poincaré generators, only $3$ are dynamical ($H,J^{x-},J^{\bar{x}-}$), compared to $4$ in the usual equal-time quantisation ($H^0,J^{0a}$).}
\begin{align}\label{hamiltonian_P2}
    H=&\,H_2+\sum_n\int d^{3n}q\;\delta\Big(\sum_i q_i\Big)h_n\,\phi^{\lambda_1}_{q_1}\cdots\phi^{\lambda_n}_{q_n}\,,\\\label{boostz_P2}
    J^{x-}=&\,J_2^{x-}+\sum_n\int d^{3n}q\;\delta\Big(\sum_i q_i\Big)\Big[j_n\,-\frac{1}{n}\,h_n\,\Big(\sum_j\frac{\partial}{\partial \bar{q}_j}\Big)\Big]\phi^{\lambda_1}_{q_1}\cdots\phi^{\lambda_n}_{q_n}\,,\\ 
    J^{\bar{x}-}=&\,J_2^{\bar{x}-}+\sum_n\int d^{3n}q\;\delta\Big(\sum_i q_i\Big)\Big[\bar{j}_n\,-\frac{1}{n}\,h_n\,\Big(\sum_j\frac{\partial}{\partial q_j}\Big)\Big]\phi^{\lambda_1}_{q_1}\cdots\phi^{\lambda_n}_{q_n}\,,
\end{align}
where we used the shorthand notation
\begin{align}
    &h_n\equiv h^{q_1,...,q_n}_{\lambda_1,...,\lambda_n}\,,&
    &j_n\equiv j^{q_1,...,q_n}_{\lambda_1,...,\lambda_n}\,,&
    &\bar{j}_n\equiv \bar{j}^{q_1,...,q_n}_{\lambda_1,...,\lambda_n}\,.
\end{align}
Let us define the momentum combinations
\begin{align}
    &\PP_{ij}=q_i\beta_j-q_j\beta_i\,,&
    &\PPb_{ij}=\bar{q}_i\beta_j-\bar{q}_j\beta_i\,,
\end{align}
where $\PPb_{ij}=-\PPb_{ji}$ and $\PP_{ij}=-\PP_{ji}$. Solving all cubic constraints required by the closure of the Poincaré algebra \cite{Bengtsson:1983pg,Bengtsson:1983pd,Bengtsson:1986kh,Metsaev:1991mt,Metsaev:1991nb} leads to the classification of cubic vertices:
\begin{align}\label{cubic_hamiltonian}
h_3=&\,C^{\lambda_1,\lambda_2,\lambda_3}\frac{\PPb^{\lambda_{123}}}{\beta_1^{\lambda_1}\beta_2^{\lambda_2}\beta_3^{\lambda_3}}+\bar{C}^{-\lambda_1,-\lambda_2,-\lambda_3}\frac{\PP^{-\lambda_{123}}}{\beta_1^{-\lambda_1}\beta_2^{-\lambda_2}\beta_3^{-\lambda_3}}\,,\\
    j_3=&\,\frac{2}{3}\,C^{\lambda_1,\lambda_2,\lambda_3}\frac{\PPb^{\lambda_{123}-1}}{\beta_1^{\lambda_1}\beta_2^{\lambda_2}\beta_3^{\lambda_3}}\Lambda^{\lambda_1,\lambda_2,\lambda_3}\,,\\
    \bar{j}_3=&\,-\frac{2}{3}\,\bar{C}^{-\lambda_1,-\lambda_2,-\lambda_3}\frac{\PP^{-\lambda_{123}-1}}{\beta_1^{-\lambda_1}\beta_2^{-\lambda_2}\beta_3^{-\lambda_3}}\Lambda^{\lambda_1,\lambda_2,\lambda_3}\,,
\end{align}
with $\lambda_{123}=\lambda_1+\lambda_2+\lambda_3$ and where we defined
\begin{align}\label{PP_cyclic}
    &\PP^a_{12}=\PP^a_{23}=\PP^a_{31}=\PP^a=\frac{1}{3}\,\Big[(\beta_1-\beta_2)q_3^a+(\beta_2-\beta_3)q_1^a+(\beta_3-\beta_1)q_2^a\Big]\,,\\
    &\Lambda^{\lambda_1,\lambda_2,\lambda_3}=\,\beta_1(\lambda_2-\lambda_3)+\beta_2(\lambda_3-\lambda_1)+\beta_3(\lambda_1-\lambda_2)\,.
\end{align}
Eq.~\eqref{PP_cyclic} follows from momentum conservation, which also implies the cyclic invariance of $\PP$ and $\PPb$: $\sigma_{123}\PP=\PP$, $\sigma_{123}\PPb=\PPb$. These expressions can be used to construct the Hamiltonian $P^-=H$ and the dynamical boost generators $J^{x-}$ and $J^{\bar{x}-}$ via Eqs.~\eqref{hamiltonian_P2} and \eqref{boostz_P2}.

Written in this form, the cubic vertices exhibit a clear separation between holomorphic and anti-holomorphic components, corresponding respectively to the terms involving $\PPb$ and $\PP$ in the equations above. This can be reached through field redefinitions, which at the cubic order correspond to the freedom of adding terms proportional to powers of the free Hamiltonian $H_2\sim \PP\PPb$, as explained in \cite{Metsaev:2005ar,Ponomarev:2016lrm}. 

Each cubic vertex in \eqref{cubic_hamiltonian} comes with an independent coupling constant, $C^{\lambda_1,\lambda_2,\lambda_3}$ or $\bar{C}^{-\lambda_1,-\lambda_2,-\lambda_3}$, which remains unfixed by the deformation procedure at the cubic level. To uncover relations among these couplings and constrain the spectrum of the theory, one must analyse the quartic consistency conditions (at the very least). In particular, it turns out that only a specific subset of these conditions --- namely, the (anti-)holomorphic constraint --- is sufficient for this purpose. On the other hand, if we are interested in theories that do not close at the cubic level and thus require a genuine quartic interaction, the full set of quartic constraints needs to be studied.

We stress that in Lorentzian signature, holomorphic and anti-holomorphic vertices are related by complex conjugation. Consequently, the reality of the action --- and hence unitarity --- requires including both sectors with cubic couplings satisfying $C^{\lambda_1,\lambda_2,\lambda_3}=(\bar{C}^{-\lambda_1,-\lambda_2,-\lambda_3})^*$. In contrast, parity invariance would require $C^{\lambda_1,\lambda_2,\lambda_3}=\bar{C}^{-\lambda_1,-\lambda_2,-\lambda_3}$. In what follows, we focus on the holomorphic sector alone, in which case the action is complex.

It is important to note that the cubic constraints, which determine the structure of the cubic vertices in \eqref{cubic_hamiltonian}, treat all cubic vertices as independent. This allows us to extend the field content by assigning to them an additional index, indicating that they belong to some representation of a gauge group $G$ with structure constants $f_{abc}$. Possible choices for $G$ include $G=U(N),SO(N)$ and $USp(N)$; for further details, see \cite{Metsaev:1991nb,Skvortsov:2020wtf,Serrani:2025owx}. One can assume that a field of helicity $\lambda$ takes values in a representation $V_\lambda$ of $G$. We will loosely denote the tensor that specifies a cubic coupling $\fA_{abc}^{\lambda_1,\lambda_2,\lambda_3}$, i.e. use the same Latin indices for all modules $V_\lambda$. Therefore, with a gauging turned on, the most general form of the cubic vertices is
\begin{equation}\label{cubic_vertex_general}
H_3^{\lambda_1,\lambda_2,\lambda_3}=\fA_{abc}^{\lambda_1,\lambda_2,\lambda_3}\int d^9q\;\delta\Big(\sum_i q_i\Big)\frac{\PPb^{\lambda_{123}}}{\beta_1^{\lambda_1}\beta_2^{\lambda_2}\beta_3^{\lambda_3}}(\phi^{\lambda_1}_{q_1})^a(\phi^{\lambda_2}_{q_2})^b(\phi^{\lambda_3}_{q_3})^c\,.
\end{equation}
It is convenient to sum over all $\lambda_{1,2,3}$ instead of all distinct triplets $\lambda_{1,2,3}$. To do so correctly, we need to impose a specific symmetry on $\fA_{abc}^{\lambda_1,\lambda_2,\lambda_3}$ for the coupling to be non-zero. Following \cite{Serrani:2025owx}, given any permutation $\sigma\in\Sigma_3$, we assume
\begin{equation}\label{coupling_sym}
    \fA_{a_{\sigma_1}a_{\sigma_2}a_{\sigma_3}}^{\lambda_{\sigma_1},\lambda_{\sigma_2},\lambda_{\sigma_3}}=(-)^{\lambda_{123}}\fA_{a_1 a_2 a_3}^{\lambda_1,\lambda_2,\lambda_3}\,.
\end{equation}
This will impose a symmetry property on the cubic couplings, but only in the case of identical fields, such as 
\begin{align}
    &\fA_{a_2a_1a_3}^{\lambda,\lambda,s}=(-)^{2\lambda+s}\fA_{a_1 a_2 a_3}^{\lambda,\lambda,s}&
    &\implies&
    &\fA_{a_2a_1a_3}=(-)^{2\lambda+s}\fA_{a_1 a_2 a_3}\,.
\end{align}
In contrast, when the fields do not carry any group indices, we must always impose a specific symmetry on the coupling constants $C^{\lambda_1,\lambda_2,\lambda_3}$. In particular, we require
\begin{equation}\label{nocolour_coupling_sym_P2}
    C^{\lambda_1,\lambda_2,\lambda_3}=(-)^{\lambda_{123}}C^{\lambda_{\sigma_1},\lambda_{\sigma_2},\lambda_{\sigma_3}}\,,
\end{equation}
where $\sigma\in\Sigma_3$. Therefore, we can assume that the coupling constants are symmetric for even-derivative interactions and antisymmetric for odd-derivative ones. This observation also implies that odd-derivative couplings involving at least two identical fields vanish by symmetry  $C^{\lambda,\lambda,\lambda'}\equiv 0$. To further simplify calculations, we can sometimes introduce generators $T_a^\lambda$ such that  $\phi^\lambda\equiv \phi^\lambda_a T^a_\lambda$ and the generators can depend on the helicity, which we often omit. This does not, of course, give the most general coupling tensor $\fA_{abc}^{\lambda_1,\lambda_2,\lambda_3}$.

Assuming the symmetry above, we can take the following form for the generators:
\begin{align}
    H_3&=\sum_{\lambda_1,\lambda_2,\lambda_3}\fA_{abc}^{\lambda_1,\lambda_2,\lambda_3}\int d^9 q\;\delta\Big(\sum_i q_i\Big)h_3\,(\phi^{\lambda_1}_{q_1})^{a}(\phi^{\lambda_2}_{q_2})^{b}(\phi^{\lambda_3}_{q_3})^{c}\,,\\
    J^{x-}_3&=\sum_{\lambda_1,\lambda_2,\lambda_3}\fA_{abc}^{\lambda_1,\lambda_2,\lambda_3}\int d^9q\;\delta\Big(\sum_i q_i\Big)\Big[j_3\,-\frac{1}{3}\,h_3\,\Big(\sum_j\frac{\partial}{\partial \bar{q}_j}\Big)\Big](\phi^{\lambda_1}_{q_1})^{a}(\phi^{\lambda_2}_{q_2})^{b}(\phi^{\lambda_3}_{q_3})^{c}\,,
\end{align}
and for the case with no gauge group:
\begin{align}
    H_3&=\sum_{\lambda_1,\lambda_2,\lambda_3}C^{\lambda_1,\lambda_2,\lambda_3}\int d^9 q\;\delta\Big(\sum_i q_i\Big)h_3\,\phi^{\lambda_1}_{q_1}\phi^{\lambda_2}_{q_2}\phi^{\lambda_3}_{q_3}\,,\\
    J^{x-}_3&=\sum_{\lambda_1,\lambda_2,\lambda_3}C^{\lambda_1,\lambda_2,\lambda_3}\int d^9q\;\delta\Big(\sum_i q_i\Big)\Big[j_3\,-\frac{1}{3}\,h_3\,\Big(\sum_j\frac{\partial}{\partial \bar{q}_j}\Big)\Big]\phi^{\lambda_1}_{q_1}\phi^{\lambda_2}_{q_2}\phi^{\lambda_3}_{q_3}\,,
\end{align}
i.e. to sum over all triplets of helicities instead of all distinct (up to permutation) triplets. 

\paragraph{Quartic consistency.} The quartic dynamical constraint takes the form
\begin{align}
    &[H,J^{a-}]\Big|_4=[H_4,J_2^{a-}]+[H_3,J_3^{a-}]-[J_4^{a-},H_2]\,,&
    &a=\{x,\bar{x}\}\,.
\end{align}
We can observe, by examining the degree of homogeneity of $q$, that the following two conditions must be satisfied independently \cite{Metsaev:1991mt,Metsaev:1991nb}
\begin{align}
    &[H_3(\PPb),J^{x-}_3]=0,&
    &[H_3(\PP),J^{\bar{x}-}_3]=0\,.
\end{align}
We refer to them, respectively, as \textit{holomorphic} and \textit{anti-holomorphic} quartic constraints. Explicitly, first the Poisson bracket between two cubic couplings is computed, then we substitute the explicit form of the generators, and once the dust has settled, we obtain
\begin{align}\label{holo_commuting_fields}
    \begin{split}
    [H_3,J_3^{x-}]=&\sum_{\lambda_i,\omega}\int d^{12}q\;\delta \left(\sum_i q_i\right)\frac{9}{2}\Big[(-)^{\omega}\frac{(\lambda_1+\omega-\lambda_2)\beta_1-(\lambda_2+\omega-\lambda_1)\beta_2}{(\beta_1+\beta_2)\beta_1^{\lambda_1}\beta_2^{\lambda_2}\beta_3^{\lambda_3}\beta_4^{\lambda_4}}\,\times\\
    &C^{\lambda_1,\lambda_2,\omega}C^{-\omega,\lambda_3,\lambda_4}\PPb_{12}^{\lambda_{12}+\omega-1}\PPb_{34}^{\lambda_{34}-\omega}\,\phi^{\lambda_1}_{q_1}\phi^{\lambda_2}_{q_2}\phi^{\lambda_3}_{q_3}\phi^{\lambda_4}_{q_4}\Big]\,,
    \end{split}
\end{align}
where $\lambda_{ij}\equiv\lambda_i+\lambda_j$. Recall that the couplings $C^{\lambda_1,\lambda_2,\lambda_3}$ obey the symmetry property \eqref{nocolour_coupling_sym_P2}. In the presence of a gauge group, we obtain
\begin{align}\label{holo_gauge_group}
    \begin{split}
    [H_3,J_3^{x-}]=&\sum_{\lambda_i,\omega}\int d^{12}q\;\delta \left(\sum_i q_i\right)\frac{9}{2}\Big[(-)^{\omega}\frac{(\lambda_1+\omega-\lambda_2)\beta_1-(\lambda_2+\omega-\lambda_1)\beta_2}{(\beta_1+\beta_2)\beta_1^{\lambda_1}\beta_2^{\lambda_2}\beta_3^{\lambda_3}\beta_4^{\lambda_4}}\,\times\\
    &\fA_{a_1a_2c}\fA^c_{\phantom{c}a_3a_4}C^{\lambda_1,\lambda_2,\omega}C^{-\omega,\lambda_3,\lambda_4}\PPb_{12}^{\lambda_{12}+\omega-1}\PPb_{34}^{\lambda_{34}-\omega}\,(\phi^{\lambda_1}_{q_1})^{a_1}(\phi^{\lambda_2}_{q_2})^{a_2}(\phi^{\lambda_3}_{q_3})^{a_3}(\phi^{\lambda_4}_{q_4})^{a_4}\Big]\,.
    \end{split}
\end{align}
In particular, it is worth noting that if we assume the theory to be purely (anti-)holomorphic and to satisfy the (anti-)holomorphic quartic constraints, it will be a well-defined theory at any order and will contain only cubic interactions.\footnote{We stress once again the chronological development of these results. In 1983, Bengtsson, Bengtsson, and Brink \cite{Bengtsson:1983pd,Bengtsson:1983pg} determined the form of the cubic interactions by solving the light-cone constraints at cubic order, and the complete classification of cubic vertices was later provided in \cite{Bengtsson:1986kh}. Several years later, in 1991, Metsaev began the analysis of the quartic constraints in \cite{Metsaev:1991mt, Metsaev:1991nb}. He observed that a simpler condition --- the holomorphic quartic constraint --- was already sufficient to fix all the cubic couplings (under certain assumptions), leading to what we call the Metsaev solution. Twenty-five years later, Ponomarev and Skvortsov revisited the problem in \cite{Ponomarev:2016lrm}. Remarkably, they found that if the theory is truncated to the (anti-)holomorphic sector, it becomes a consistent theory with only cubic vertices at all orders, and the couplings are the same as those found by Metsaev. More recently, in \cite{Serrani:2025owx}, it was shown that the holomorphic constraint admits multiple consistent truncations, even allowing for theories with a finite spectrum that still involve higher-spin fields.} 

\section{OPE associativity in 2d CCFT}\label{Paper2_section3}
Recently, considerable attention has been devoted to the study of the \textit{OPE associativity constraint}, see \cite{Mago:2021wje,Ren:2022sws,Monteiro:2022lwm,Costello:2022upu,Bittleston:2022jeq,Ball:2022bgg,Ball:2023sdz,Ball:2023qim,Ball:2024oqa,Fernandez:2024qnu,Guevara:2024ixn,Bhattacharyya:2025nfp}. Here, we briefly review the basics of $2d$ CCFT for massless particles (see Appendix \ref{Paper2_AppendixB} for notations), to provide context for the OPE associativity constraint. For more details and reviews on the topic, see \cite{Pasterski:2021rjz,Raclariu:2021zjz}.

In CCFT, of fundamental importance is the isomorphism $SO^+(1,3)\simeq SL(2,\mathbb{C})/\mathbb{Z}_2$, between the $4d$ (connected) Lorentz group and the set of conformal transformations of the $2d$ celestial sphere $\mathbb{C}\PP^1$\cite{oblak2018lorentzgroupcelestialsphere}. 

This is what allows the S-matrix in asymptotically flat spacetime to be recast as a celestial amplitude. The idea is to perform a change of basis: from the standard energy-momentum eigenstate basis, with definite energy and momentum, to a new basis of boost eigenstates. In this new basis, the S-matrix is interpreted as a correlation function of operators inserted at points on the celestial sphere at null infinity. These operators are labelled by the conformal dimension and spin associated with the $2d$ global conformal group of the celestial sphere $SL(2,\mathbb{C})/\mathbb{Z}_2$.

The $S$-matrix element for the scattering of $n$ massless fields in the standard energy-momentum eigenstate basis is 
\begin{equation}
    \mathcal{A}_n(\omega_j,z_j,\bar{z}_j)=\langle \text{out}|S|\text{in}\rangle\,.
\end{equation}
The integral operation that allows one to perform the change of basis to the boost eigenstate one and then trades the energy\footnote{To be precise, the energy is given by $k^0=\omega q^0=\omega(1+z\bar{z})$.} $\omega$ for the conformal dimension $\Delta$ is the Mellin transform
\begin{equation}
    M(\cdot)=\int_0^{\infty}\frac{d\omega}{\omega}\omega^{\Delta}(\cdot)\,.
\end{equation}
Then the $S$-matrix in the boost eigenstates (i.e. the celestial amplitude) is given by the Mellin-transformed amplitude
\begin{equation}\label{Mellin_tranform_amplitude}
\mathcal{M}_n(\Delta_j,z_j,\bar{z}_j)=\prod_{j=1}^n\int_0^{\infty}\frac{d\omega_j}{\omega_j}\omega_j^{\Delta_j}\mathcal{A}_n(\omega_j,z_j,\bar{z}_j)\,,
\end{equation}
where, by definition, we have
\begin{equation}
    \mathcal{M}_n(\Delta_j,z_j,\bar{z}_j)\equiv \prescript{}{boost}{\langle} \text{out}|S|\text{in}\rangle_{boost}=\langle\mathcal{O}^{\pm}_{\Delta_1,s_1}(z_1,\bar{z}_1)\cdots \mathcal{O}^{\pm}_{\Delta_n,s_n}(z_n,\bar{z}_n)\rangle_{CCFT}\,,
\end{equation} 
where the operator $\mathcal{O}^{\pm}_{\Delta_n,s_n}$ represents outgoing $(+)$ or incoming $(-)$ celestial conformal primary operators with $2d$ conformal (or boost) weight $\Delta_n$ and spin $s_n$ that crosses the celestial sphere at a point $(z,\bar{z})$. In particular \cite{Pasterski:2017kqt}, when $\Delta_n\in 1+i\mathbb{R}$, corresponding to the principal continuous series of the Lorentz group, the transformation above can be inverted using the inverse Mellin transform 
\begin{equation}
    \mathcal{A}_n(\omega_j,z_j,\bar{z}_j)=\prod_{i=1}^n\int_{1-i\infty}^{1+i\infty}\frac{d\Delta_j}{2\pi i}\omega_j^{-\Delta_i}\mathcal{M}_n(\Delta_j,z_j,\bar{z}_j)\,.
\end{equation}
From this point on, we omit the $\pm$ sign and adopt the all–outgoing convention, as in Appendix \ref{Paper2_AppendixB}.

One of the central ingredients in any conformal field theory is the operator product expansion (OPE), which provides a systematic way to compute products of local operators and correlation functions. The holomorphic OPE of two conformal primary operators with conformal weights $(h_i,\bar{h}_i)$, where $s_i=h_i-\bar{h}_i$ and $\Delta_i=h_i+\bar{h}_i$, in celestial CFT can be obtained by the holomorphic collinear limit (i.e. $z_1\rightarrow z_2$) of scattering amplitudes \cite{Fan:2019emx,Pate:2019lpp,Himwich:2021dau} and can be expressed as
\begin{align}\label{CelestialOPE}
   & \mathcal{O}_{h_1,\bar{h}_1}(z_1,\bar{z}_1)\mathcal{O}_{h_2,\bar{h}_2}(z_2,\bar{z}_2)\sim\frac{1}{z_{12}}\sum_p\sum_{m=0}^{\infty}C_p^{(m)}(\bar{h}_1,\bar{h}_2)\bar{z}_{12}^{p+m}\bar{\partial}^m\mathcal{O}_{h_{12}-1,\bar{h}_{12}+p}(z_2,\bar{z}_2)\,,\\
    & z_{ij}=z_i-z_j\,,\quad
    \bar{z}_{ij}=\bar{z}_i-\bar{z}_j\,,\quad
    h_{ij}=h_i+h_j\,,\quad
    \bar{h}_{ij}=\bar{h}_i+\bar{h}_j\,.\quad
    s_{ij}=s_i+s_j\,,
\end{align}
where $C_p^{(m)}(\bar{h}_1,\bar{h}_2)$ denotes the OPE coefficient determining the contribution of the $m$th right-moving descendant with weights $(h_{12}-1,\bar{h}_{12}+p+m)$ and is given by
\begin{equation}
    C_p^{(m)}(\bar{h}_1,\bar{h}_2)=-\frac{1}{2}C^{s_1,s_2,s_3}\frac{1}{m!}B(2\bar{h}_1+p+m,2\bar{h}_2+p)\,,
\end{equation}
where $B(a,b)$ is the Euler beta function and $C^{s_1,s_2,s_3}$ is the coupling constant appearing in the flat bulk $3$-pt function of massless (higher-spin) particles with spins $(s_1,s_2,s_3=p+1-s_{12})$. Therefore, there is a one-to-one correspondence between holomorphic bulk $3$-pt vertices and holomorphic celestial OPE.

Notice that the OPE expansion \eqref{CelestialOPE} represents only a partial result, as it captures solely the holomorphic sector at tree level. In fact, one-loop contributions already introduce double poles $\sim \frac{1}{z_{12}^2}$, arising from massless loops, as well as logarithms such as $\sim\log z_{12}$ \cite{Costello:2022upu, Bittleston:2022jeq, Ball:2023qim, Bhardwaj:2022anh, Krishna:2023ukw, Bhardwaj:2024wld, Bissi:2024brf}. Moreover, taking into account the anti-holomorphic sector leads to a more complete, ``all-order'' celestial OPE that involves series expansions in powers of both $\bar{z}$ and $z$. This has so far been achieved only in the maximally helicity-violating (MHV) sector at tree level, via twistor string theory in \cite{Adamo:2022wjo} and through collinear expansions using on-shell recursion relations in \cite{Ren:2023trv}.

A crucial consistency condition --- expected to be necessary for the existence of a well-defined celestial dual theory of gravity --- is the associativity of the OPE. We now review the (anti-)holomorphic OPE associativity constraint in $2d$ CCFT as derived in \cite{Ren:2022sws}, which implies the Jacobi identity for the charges studied in \cite{Mago:2021wje}.

A standard way to check OPE associativity is to introduce a mode expansion of the operators involved and compute the commutators using their OPE, see \cite{Guevara:2021abz}. Then OPE associativity implies the Jacobi identity for the modes. This is of particular interest when applied to conformally soft currents, see \cite{Mago:2021wje}. The relation is made possible thanks to the commutators for holomorphic objects \cite{Raclariu:2021zjz,Strominger:2021mtt} determined by a contour integral as
\begin{equation}
    [A,B](z_1)=\oint_{z_1}\frac{d z_2}{2\pi i}A(z_2)B(z_1)=\underset{z_2\rightarrow z_1}{\text{Res}}A(z_2)B(z_1)\,.
\end{equation}
We start with the following identity between correlators of conformal primaries involving contour integrals\footnote{This was the method used in \cite{Ren:2022sws}. A related one–loop analysis for self-dual Yang–Mills was already carried out in \cite{Costello:2022upu}, where extra powers of $z_k$ in the Jacobi identity are needed to extract higher–order poles coming from the one-loop OPE. At tree level, no such terms are required.}  
\begin{align}
    \begin{split}
    \Bigg(\oint_{|z_{13}|=2}dz_1\oint_{|z_{23}|=1}dz_2\,-&\oint_{|z_{23}|=2}dz_2\oint_{|z_{13}|=1}dz_1\,-\oint_{|z_{23}|=2}dz_2\oint_{|z_{12}|=1}dz_1\Bigg)\times\\
    &\langle\mathcal{O}^{a_1}_{\Delta_1,s_1}(z_1,\bar{z}_1)\cdots \mathcal{O}^{a_n}_{\Delta_n,s_n}(z_n,\bar{z}_n)\rangle=0\,,
    \end{split}
\end{align}
where we allow the presence of internal indices, enabling the operators to live in specific representations of an internal gauge group. This can be rewritten in terms of a ``double residue condition'' as
\begin{equation}\label{OPE_associativity}
\left(\underset{z_1\rightarrow z_3}{\text{Res}}\,\underset{z_2\rightarrow z_3}{\text{Res}}-\underset{z_2\rightarrow z_3}{\text{Res}}\,\underset{z_1\rightarrow z_3}{\text{Res}}-\underset{z_2\rightarrow z_3}{\text{Res}}\,\underset{z_1\rightarrow z_2}{\text{Res}}\right)\langle\mathcal{O}^{a_1}_{\Delta_1,s_1}(z_1,\bar{z}_1)\cdots \mathcal{O}^{a_n}_{\Delta_n,s_n}(z_n,\bar{z}_n)\rangle=0\,.
\end{equation}
For further discussion of this constraint and possible ambiguities in its definition, see \cite{Ball:2022bgg,Ball:2024oqa}. To express \eqref{OPE_associativity} in terms of the standard momentum space amplitude, we use the Mellin transform \eqref{Mellin_tranform_amplitude} and compute the residues of the amplitude $\mathcal{A}_n(\omega_j,z_j,\bar{z}_j)$ as
\begin{equation}\label{OPE_amplitude}
\left(\underset{z_1\rightarrow z_3}{\text{Res}}\,\underset{z_2\rightarrow z_3}{\text{Res}}-\underset{z_2\rightarrow z_3}{\text{Res}}\,\underset{z_1\rightarrow z_3}{\text{Res}}-\underset{z_2\rightarrow z_3}{\text{Res}}\,\underset{z_1\rightarrow z_2}{\text{Res}}\right)\mathcal{A}_n(1\cdots n)=0\,.
\end{equation}
At the amplitude level, the residue can be efficiently computed by taking the appropriate collinear limits. Schematically, proceeding step by step, we first take the collinear limit $q_1^{\mu}\parallel q_2^{\mu}$, which gives
\begin{equation}
    \lim_{z_{12}\rightarrow 0}\mathcal{A}_n(12\cdots n)=\sum_{\lambda_i}\frac{\mathcal{A}_3(12i)\mathcal{A}_{n-1}(i34\cdots n)}{\langle 12\rangle [12]}\,,
\end{equation}
where we focus on the leading order in $\frac{1}{z_{12}}$ and possible other channels decouple in the limit. Then it is followed by the collinear limit $q_2^{\mu}\parallel q_3^{\mu}$, and we get
\begin{equation}
    \lim_{z_{23}\rightarrow 0}\,\lim_{z_{12}\rightarrow 0}\mathcal{A}_n(12\cdots n)=\sum_{\lambda_i}\sum_{\lambda_j}\frac{\mathcal{A}_3(12i)\mathcal{A}_3(i3j)\mathcal{A}_{n-2}(j4\cdots n)}{\langle 12\rangle [12]\langle i3\rangle [i3]}\,.
\end{equation}
Then we can extract the residue of the last expression as
\begin{align}\label{computing_residues}
    \begin{split}
    &\underset{z_2\rightarrow z_3}{\text{Res}}\,\underset{z_1\rightarrow z_2}{\text{Res}}\mathcal{A}_n(12\cdots n)=\sum_{\lambda_i}\sum_{\lambda_j}\frac{\mathcal{A}_3(12i)\mathcal{A}_3(i3j)\mathcal{A}_{n-2}(j4\cdots n)}{\omega_1\omega_2\omega_i\omega_3 \bar{z}_{12} \bar{z}_{i3}}\,,\\
    &\bar{z}_i=\frac{\omega_1\bar{z}_1+\omega_2\bar{z}_2}{\omega_1+\omega_2}\,,\qquad
    \omega_i=\omega_1+\omega_2\,,
    \end{split}
\end{align}
where we used the notation introduced in Appendix \ref{Paper2_AppendixC}. Using the expression for the double residue \eqref{computing_residues} inside \eqref{OPE_amplitude}, we get\footnote{From now on, we change notation and denote the spin $s_n$ of the massless fields by the helicity $\lambda_n$, in line with the light-cone higher-spin literature.}
\begin{align}\label{OPE_Ass_commuting}
    \begin{split}
\sum_{\lambda_i}\Bigg[&\frac{\mathcal{A}_3(p_1^{\lambda_1},p_2^{\lambda_2},p_i^{\lambda_i})\mathcal{A}_3(p_3^{\lambda_3},p_4^{\lambda_4},p_i^{-\lambda_i})}{\bar{z}_{12}\bar{z}_{i3}(\omega_1+\omega_2)\omega_1\omega_2\omega_3}+\frac{\mathcal{A}_3(p_2^{\lambda_2},p_3^{\lambda_3},p_i^{\lambda_i})\mathcal{A}_3(p_1^{\lambda_1},p_4^{\lambda_4},p_i^{-\lambda_i})}{\bar{z}_{23}\bar{z}_{i1}(\omega_2+\omega_3)\omega_1\omega_2\omega_3}\\
    &+\frac{\mathcal{A}_3(p_3^{\lambda_3},p_1^{\lambda_1},p_i^{\lambda_i})\mathcal{A}_3(p_2^{\lambda_2},p_4^{\lambda_4},p_i^{-\lambda_i})}{\bar{z}_{31}\bar{z}_{i2}(\omega_1+\omega_3)\omega_1\omega_2\omega_3}\Bigg]=0\,,
    \end{split}
\end{align}
where we dropped the sum over $\lambda_j$ because the OPE must vanish for a generic $\mathcal{A}_{n-2}$, so the constraint must hold separately for each $\lambda_j$. For simplicity, we relabel $\lambda_j$ as $\lambda_4$.

If we consider fields transforming in a matrix representation of a Lie algebra, the associated generators $T^{a_n}$ are carried along throughout the computation. As shown above, the procedure remains essentially the same, with the only difference being that we can factor out the trace over the generators. The clearest way to illustrate this is to begin with the colour amplitudes, written in terms of the colour-ordered ones as
\begin{equation}\label{colour_ordered_ampl}
    \mathcal{A}(12\cdots n)=\sum_{\sigma\in S_n/\mathbb{Z}_n}\mathrm{Tr}(T^{a_{\sigma_1}}T^{a_{\sigma_2}}\cdots T^{a_{\sigma_n}})\tilde{\mathcal{A}}(\sigma_1\sigma_2\cdots\sigma_n)\,.
\end{equation}
Consequently, the constraint \eqref{OPE_amplitude} takes the form
\begin{equation}\label{OPE_Ass_U(N)}
    \sum_{\sigma\in S_n/\mathbb{Z}_n}\sum_{\lambda_i}\Bigg[\frac{\tilde{\mathcal{A}}_3(p_{\sigma_1}^{\lambda_{\sigma_1}},p_{\sigma_2}^{\lambda_{\sigma_2}},p_{\sigma_i}^{\lambda_{\sigma_i}})\tilde{\mathcal{A}}_3(p_{\sigma_3}^{\lambda_{\sigma_3}},p_{\sigma_4}^{\lambda_{\sigma_4}},p_{\sigma_i}^{-\lambda_{\sigma_i}})}{\bar{z}_{\sigma_1\sigma_2}\bar{z}_{\sigma_i\sigma_3}(\omega_{\sigma_1}+\omega_{\sigma_2})\omega_{\sigma_1}\omega_{\sigma_2}\omega_{\sigma_3}}\mathrm{Tr}(T^{a_{\sigma_1}}T^{a_{\sigma_2}}T^{a_{\sigma_3}}T^{a_{\sigma_4}})\Bigg]=0\,.
\end{equation}
These two constraints \eqref{OPE_Ass_commuting} and \eqref{OPE_Ass_U(N)} are the celestial OPE associativity constraints, which we will solve in full generality. A similarity with the previously discussed light-cone constraint is already apparent (in the sense that both consist of two three-point vertices brought together times some additional kinematical factors), and we will aim to establish a direct connection between the two in the following sections.

Even more in general, we can assume, as considered for the holomorphic constraint, fields belonging to some representation of a gauge group $G$ with generic structure constants $f_{abc}$, and arrive at
\begin{equation}
    \sum_{\sigma\in S_n/\mathbb{Z}_n}\sum_{\lambda_i}\Bigg[\frac{\tilde{\mathcal{A}}_{a_{\sigma_1}a_{\sigma_2}}^{\phantom{a_{\sigma_1}a_{\sigma_2}}c}(p_{\sigma_1}^{\lambda_{\sigma_1}},p_{\sigma_2}^{\lambda_{\sigma_2}},p_{\sigma_i}^{\lambda_{\sigma_i}})\tilde{\mathcal{A}}_{c\,a_{\sigma_3}a_{\sigma_4}}(p_{\sigma_3}^{\lambda_{\sigma_3}},p_{\sigma_4}^{\lambda_{\sigma_4}},p_{\sigma_i}^{-\lambda_{\sigma_i}})}{\bar{z}_{\sigma_1\sigma_2}\bar{z}_{\sigma_i\sigma_3}(\omega_{\sigma_1}+\omega_{\sigma_2})\omega_{\sigma_1}\omega_{\sigma_2}\omega_{\sigma_3}}T^{a_{\sigma_1}}T^{a_{\sigma_2}}T^{a_{\sigma_3}}T^{a_{\sigma_4}}\Bigg]=0\,,
\end{equation}
where we denoted $\mathcal{A}_3=\tilde{\mathcal{A}}_{abc}T^aT^bT^c=\tilde{\mathcal{A}}_3\fA_{abc}T^aT^bT^c$. We solve this more general case in Appendix \ref{Paper2_AppendixD}.

\section{Celestial OPE associativity constraint}\label{Paper2_section4}
In this section, we present an exact solution to the OPE associativity constraint by switching to the light-front approach\footnote{We will use the terms light-front and light-cone interchangeably throughout this work.}. It turns out to be very useful to rewrite the OPE associativity constraint in terms of the light-cone variables. This can be done by following Appendix \ref{Paper2_AppendixC}, where we explain how to relate the two formalisms through the spinor-helicity one. The result is that, on-shell, we have the following identity:
\begin{equation}
    \sum_{\lambda_i}\frac{\mathcal{A}_3(p_1^{\lambda_1},p_2^{\lambda_2},p_i^{\lambda_i})\mathcal{A}_3(p_3^{\lambda_3},p_4^{\lambda_4},p_i^{-\lambda_i})}{\bar{z}_{12}\bar{z}_{i3}(\omega_1+\omega_2)\omega_1\omega_2\omega_3}=\sum_{\lambda_i}(-)^{\lambda_i}C^{\lambda_1,\lambda_2,\lambda_i}C^{-\lambda_i\lambda_3,\lambda_4}\frac{\PPb_{12}^{\lambda_{12}+\lambda_i-1}\PPb_{34}^{\lambda_{34}-\lambda_i-1}}{\beta_1^{\lambda_1}\beta_2^{\lambda_2}\beta_3^{\lambda_3}\beta_4^{\lambda_4}}\,.
\end{equation}
The right-hand side is not the same term present in the quartic holomorphic constraint, but, as we will see in the final section, the OPE associativity constraint is closely related to it. The quartic holomorphic constraint was solved in \cite{Serrani:2025owx} in two specific cases: for singlet fields and for fields living in some matrix representation of a gauge group $G$.

If we wish to reproduce all lower-spin constraints arising from the OPE associativity \eqref{OPE_Ass_commuting} and \eqref{OPE_Ass_U(N)}, as found in \cite{Ren:2022sws}, we need to consider the following types of cubic couplings:
\begin{align}\label{h3_1}
&h_3^{(1)}\sim\phi^{\lambda_1}_{q_1}\phi^{\lambda_2}_{q_2}\phi^{\lambda_3}_{q_3}\,,\\\label{h3_2}
    &h_3^{(2)}\sim\mathrm{Tr}(T_{a_1}T_{a_2})(\phi^{\lambda_1}_{q_1})^{a_1}(\phi^{\lambda_2}_{q_2})^{a_2}\phi^{\lambda_3}_{q_3}\,,\\\label{h3_3}
    &h_3^{(3)}\sim\mathrm{Tr}(T_{a_1}T_{a_2}T_{a_3})(\phi^{\lambda_1}_{q_1})^{a_1}(\phi^{\lambda_2}_{q_2})^{a_2}(\phi^{\lambda_3}_{q_3})^{a_3}\,.
\end{align}
For semisimple compact Lie algebras, we can write
\begin{align}\label{semisimple_a}
    &\mathrm{Tr}(T_{a_1}T_{a_2})=\frac{1}{2}\delta_{a_1a_2}\,,\\
    &\mathrm{Tr}(T_{a_1}T_{a_2}T_{a_3})=\frac{1}{2}\big(\mathrm{Tr}(T_{a_1}\{T_{a_2},T_{a_3}\})+\mathrm{Tr}(T_{a_1}[T_{a_2},T_{a_3}])\big)=\frac{1}{4}(d_{a_1a_2a_3}+if_{a_1a_2a_3})\,,
\end{align}
where the first line is a normalisation that can always be achieved, $f_{a_1a_2a_3}$ denotes the antisymmetric structure constant, and $d_{a_1a_2a_3}$ the symmetric $d$-symbol, both defined as
\begin{align}
    &[T_{a_1},T_{a_2}]=if_{a_1a_2a_3}T^{a_3}\,,&
    &\{T_{a_1},T_{a_2}\}=\frac{1}{N}\delta_{a_1a_2}\mathbb{I}+d_{a_1a_2a_3}T^{a_3}\,,\\\label{semisimple_b}
    &if_{a_1a_2a_3}\equiv 2\,\mathrm{Tr}(T_{a_1}[T_{a_2},T_{a_3}])\,,&
    &d_{a_1a_2a_3}\equiv 2\,\mathrm{Tr}(T_{a_1}\{T_{a_2},T_{a_3}\})\,.
\end{align}
Looking at the various couplings in \eqref{h3_1}--\eqref{h3_3}, we can identify five possible structures that can occur at the quartic order:
\begin{align}\label{normal_case_1}
&(h_3^{(1)},h_3^{(1)})&&\phi^{\lambda_1}_{q_1}\phi^{\lambda_2}_{q_2}\phi^{\lambda_3}_{q_3}\phi^{\lambda_4}_{q_4}\,,\\\label{normal_case_2}
&(h_3^{(3)},h_3^{(3)})&&\mathrm{Tr}(T_{a_1}T_{a_2}T_{a_3}T_{a_4})(\phi^{\lambda_1}_{q_1})^{a_1}(\phi^{\lambda_2}_{q_2})^{a_2}(\phi^{\lambda_3}_{q_3})^{a_3}(\phi^{\lambda_4}_{q_4})^{a_4}\,,\\\label{mixed_case_1}
&(h_3^{(2)},h_3^{(2)})\,,(h_3^{(1)},h_3^{(2)})&&\mathrm{Tr}(T_{a_1}T_{a_2})(\phi^{\lambda_1}_{q_1})^{a_1}(\phi^{\lambda_2}_{q_2})^{a_2}\phi^{\lambda_3}_{q_3}\phi^{\lambda_4}_{q_4}\,,\\\label{mixed_case_2}
&(h_3^{(2)},h_3^{(2)})&&\mathrm{Tr}(T_{a_1}T_{a_2})\mathrm{Tr}(T_{a_3}T_{a_4})(\phi^{\lambda_1}_{q_1})^{a_1}(\phi^{\lambda_2}_{q_2})^{a_2}(\phi^{\lambda_3}_{q_3})^{a_3}(\phi^{\lambda_4}_{q_4})^{a_4}\,,\\\label{mixed_case_3}
&(h_3^{(2)},h_3^{(3)})&&\mathrm{Tr}(T_{a_1}T_{a_2}T_{a_3})(\phi^{\lambda_1}_{q_1})^{a_1}(\phi^{\lambda_2}_{q_2})^{a_2}(\phi^{\lambda_3}_{q_3})^{a_3}\phi^{\lambda_4}_{q_4}\,,
\end{align}
where on the left of each structure, we indicate the type of pair of cubic vertices responsible for generating each distinct type of constraint. In particular, we observe that while the first two involve the same type of cubic vertices, the others mix and intertwine different types of cubic vertices. As we will see, there are two main mechanisms by which the constraint can be solved. Every type of constraint will fall into one of these two cases.

\subsection{Singlet fields}
We start by solving the OPE associativity constraint \cite{Ren:2022sws}, for singlet fields \eqref{normal_case_1}, then
\begin{align}\label{OPE_Ass}
    \begin{split}
\sum_{\lambda_i}\Bigg[&\frac{\mathcal{A}_3(p_1^{\lambda_1},p_2^{\lambda_2},p_i^{\lambda_i})\mathcal{A}_3(p_3^{\lambda_3},p_4^{\lambda_4},p_i^{-\lambda_i})}{\bar{z}_{12}\bar{z}_{i3}(\omega_1+\omega_2)\omega_1\omega_2\omega_3}+\frac{\mathcal{A}_3(p_2^{\lambda_2},p_3^{\lambda_3},p_i^{\lambda_i})\mathcal{A}_3(p_1^{\lambda_1},p_4^{\lambda_4},p_i^{-\lambda_i})}{\bar{z}_{23}\bar{z}_{i1}(\omega_2+\omega_3)\omega_1\omega_2\omega_3}\\
    &+\frac{\mathcal{A}_3(p_3^{\lambda_3},p_1^{\lambda_1},p_i^{\lambda_i})\mathcal{A}_3(p_2^{\lambda_2},p_4^{\lambda_4},p_i^{-\lambda_i})}{\bar{z}_{31}\bar{z}_{i2}(\omega_1+\omega_3)\omega_1\omega_2\omega_3}\Bigg]=0\,,
    \end{split}
\end{align}
and we rewrite it in the light-cone formalism (see Appendix \ref{Paper2_AppendixC}) as
\begin{align}\label{OPE_Ass_LC}
\begin{split}
\sum_{\lambda_i}(-)^{\lambda_i}\Big(&C^{\lambda_1,\lambda_2,\lambda_i}C^{-\lambda_i,\lambda_3,\lambda_4}\PPb_{12}^{\lambda_{12}+\lambda_i-1}\PPb_{34}^{\lambda_{34}-\lambda_i-1}+C^{\lambda_1,\lambda_4,\lambda_i}C^{-\lambda_i,\lambda_2,\lambda_3}\PPb_{14}^{\lambda_{14}+\lambda_i-1}\PPb_{23}^{\lambda_{23}-\lambda_i-1}\\
&+C^{\lambda_3,\lambda_1,\lambda_i}C^{-\lambda_i,\lambda_2,\lambda_4}\PPb_{31}^{\lambda_{13}+\lambda_i-1}\PPb_{24}^{\lambda_{24}-\lambda_i-1}\Big)=0\,,\qquad \lambda_{ij}=\lambda_i+\lambda_j\,.
\end{split}
\end{align}
This structure is reminiscent of the quartic holomorphic constraint studied in \cite{Ponomarev:2016lrm,Serrani:2025owx}. As we will see in the final section, the two are closely related. Moreover, our strategy to solve the constraint follows the same ideas used to address the quartic holomorphic constraint in \cite{Serrani:2025owx}. To avoid unnecessary minus signs between products of couplings and extra $i$ factors between couplings, we assign an additional factor of $i$ to odd-helicity fields. This is equivalent to taking even-helicity fields to be Hermitian and odd-helicity fields to be anti-Hermitian matrices (with singlets treated as $1\times 1$ matrices). With this convention, the factor $(-)^{\lambda_i}$ in \eqref{OPE_Ass_LC} is removed. We begin by introducing three independent variables:
\begin{align}\label{ABC_variables}
\begin{split}
    2A&=\PPb_{12}+\PPb_{34}=\PPb_{23}-\PPb_{14}\,,\qquad
    2B=\PPb_{13}-\PPb_{24}=\PPb_{34}-\PPb_{12}\,,\\
    2C&=\PPb_{14}+\PPb_{23}=-\PPb_{13}-\PPb_{24}\,,
    \end{split}
\end{align}
with the following transformation properties:
\begin{subequations}\label{ABC_sym}
\begin{align}
    &A\overset{1\leftrightarrow 2}{\xleftrightarrow{\hspace{5mm}}} B\,,&
    &A\overset{1\leftrightarrow 3}{\xleftrightarrow{\hspace{5mm}}} -A\,,&
    &A\overset{1\leftrightarrow 4}{\xleftrightarrow{\hspace{5mm}}} C\,,&
    &A\overset{2\leftrightarrow 3}{\xleftrightarrow{\hspace{5mm}}} -C\,,&
    &A\overset{2\leftrightarrow 4}{\xleftrightarrow{\hspace{5mm}}} -A\,,&
    &A\overset{3\leftrightarrow 4}{\xleftrightarrow{\hspace{5mm}}} -B\,,\\
    &C\overset{1\leftrightarrow 2}{\xleftrightarrow{\hspace{5mm}}} -C\,,&
    &B\overset{1\leftrightarrow 3}{\xleftrightarrow{\hspace{5mm}}} C\,,&
    &B\overset{1\leftrightarrow 4}{\xleftrightarrow{\hspace{5mm}}} -B\,,&
    &B\overset{2\leftrightarrow 3}{\xleftrightarrow{\hspace{5mm}}} -B\,,&
    &B\overset{2\leftrightarrow 4}{\xleftrightarrow{\hspace{5mm}}} -C\,,&
    &C\overset{3\leftrightarrow 4}{\xleftrightarrow{\hspace{5mm}}} -C\,,
\end{align}
\end{subequations}
and define
\begin{align}\label{definitions}
\nonumber
    &\mathcal{C}^{1234\lambda_i}\definition C^{\lambda_1,\lambda_2,\lambda_i}C^{-\lambda_i,\lambda_3,\lambda_4}\,,\qquad
        k_+^{1234}\definition (-)^{\lambda_{12}}\sum_{\lambda_i}(-)^{\lambda_i}\mathcal{C}^{1234\lambda_i}\,,\qquad
        k_-^{1234}\definition \sum_{\lambda_i}\mathcal{C}^{1234\lambda_i}\,,\\
    &f_+^{1234}(A,B)\definition (-)^{\lambda_{34}}\sum_{\lambda_i}(-)^{\lambda_i}\mathcal{C}^{1234\lambda_i}(A-B)^{\lambda_{12}+\lambda_i-1}(A+B)^{\lambda_{34}-\lambda_i-1}\,,\\
    \nonumber
    &f_-^{1234}(A,B)\definition \sum_{\lambda_i}\mathcal{C}^{1234\lambda_i}(A-B)^{\lambda_{12}+\lambda_i-1}(A+B)^{\lambda_{34}-\lambda_i-1}\,,
\end{align}
where the upper index $1234$ denotes the order of the external helicities. We note immediately that $f_-^{1234}(A,B)=f_+^{1234}(B,A)(-)^{\Lambda-1}$, where $\Lambda=\lambda_1+\lambda_2+\lambda_3+\lambda_4$, but it is convenient to keep both of them. The OPE associativity constraint can be recast as
\begin{align}\label{AssOPE}
\nonumber
\sum_{\lambda_i}&\Big(\mathcal{C}^{1234\lambda_i}(A-B)^{\lambda_{12}+\lambda_i-1}(A+B)^{\lambda_{34}-\lambda_i-1}+(-)^{\Lambda}\mathcal{C}^{3124\lambda_i}(B-C)^{\lambda_{13}+\lambda_i-1}(B+C)^{\lambda_{24}-\lambda_i-1}\\
&+\mathcal{C}^{1423\lambda_i}(C-A)^{\lambda_{14}+\lambda_i-1}(C+A)^{\lambda_{23}-\lambda_i-1}\Big)=0\,,
\end{align}
and plugging in the definitions \eqref{definitions} for the functions $f^{\cdot\cdots}_-$, we have
\begin{equation}\label{AssOPEv2}
f^{1234}_-(A,B)+(-)^{\Lambda}f^{3124}_-(B,C)+f^{1423}_-(C,A)=0\,.
\end{equation}
The most general polynomial form of the functions $f^{\cdot\cdots}_-$ that is compatible with the constraint \eqref{AssOPEv2}, is the following:\footnote{For instance, choosing $f^{1234}_-(A,B)\sim A^iB^j$, with $i,j>0$ and $i+j=\Lambda-2$, would never satisfy \eqref{AssOPEv2}, since neither $f^{3124}_-(B,C)$ nor $f^{1423}_-(C,A)$ could cancel such contributions.}
\begin{subequations}\label{eq_all}
\begin{align}
    f^{1234}_-(A,B)&=k^{1234}_-A^{\Lambda-2}-k^{1234}_+B^{\Lambda-2}\,,\\
    f^{3124}_-(B,C)&=k^{3124}_-B^{\Lambda-2}-k^{3124}_+C^{\Lambda-2}\,,\\
    f^{1423}_-(C,A)&=k^{1423}_-C^{\Lambda-2}-k^{1423}_+A^{\Lambda-2}\,.
\end{align}
\end{subequations}
where $k^{\cdots\cdot}$ are some free coefficients (that we identify with those in \eqref{definitions}).
Substituting these back into \eqref{AssOPEv2} gives the constraints
\begin{align}\label{condition_eq1}
    &k^{1234}_-=k^{1423}_+\,,&
    &k^{3124}_-=(-)^{\Lambda}k^{1234}_+\,,&
    &k^{1423}_-=(-)^{\Lambda}k^{3124}_+\,.
\end{align}
The form of the functions in \eqref{eq_all} plus the conditions in \eqref{condition_eq1} gives the solution to the OPE associativity constraint \eqref{AssOPEv2}. To express the solution in terms of the cubic couplings, we need to solve \eqref{eq_all} in terms of the product of couplings $\mathcal{C}^{\bullet\bullet\bullet\bullet\bullet}$. This can be done by expressing \eqref{eq_all} using the previous variables $\PP_{34}=A+B$ and $\PP_{12}=A-B$, as follows
\begin{equation}
    \sum_{\lambda_i}\mathcal{C}^{1234\lambda_i}\PP_{12}^{\lambda_{12}+\lambda_i-1}\PP_{34}^{\lambda_{34}-\lambda_i-1}=k^{1234}_-\left(\frac{\PP_{34}+\PP_{12}}{2}\right)^{\Lambda-2}-k^{1234}_+\left(\frac{\PP_{34}-\PP_{12}}{2}\right)^{\Lambda-2}\,.
\end{equation}
By expanding the Binomial on the RHS and matching the monomials in terms of $\PP_{34}$ and $\PP_{12}$ we get
\begin{equation}\label{Ass_system_eq}
    \mathcal{C}^{1234\lambda_i}=\frac{(k^{1234}_-+(-)^{\lambda_i+\lambda_{12}}k^{1234}_+)(\Lambda-2)!}{2^{\Lambda-2}(\lambda_{12}+\lambda_i-1)!(\lambda_{34}-\lambda_i-1)!}\quad
    \forall\;\lambda_i\,,\quad \text{same for $(3124)$ and $(1423)$}\,.\\
\end{equation}
Summing over $\lambda_i$, we can check consistency:
\begin{equation}\label{consistency_1}
   \sum_{\lambda_i=-\lambda_{12}+1}^{\lambda_{34}-1}\frac{(\Lambda-2)!}{2^{\Lambda-2}(\lambda_{12}+\lambda_i-1)!(\lambda_{34}-\lambda_i-1)!}=\frac{1}{2^{\Lambda-2}}\sum_{n=0}^{\Lambda-2}
   \begin{pmatrix}
       \Lambda-2\\
       n
   \end{pmatrix}
   =1\,.
\end{equation}
Notice that, if we consider only even-derivative vertices (i.e. with $k^{\cdots\cdot}_-=k^{\cdots\cdot}_+$), we get
\begin{equation}
    k^{1234}_-=k^{3124}_-=k^{1423}_-=k^{1234}_+=k^{3124}_+=k^{1423}_+\,.
\end{equation}
This case makes the classification of solutions to the OPE associativity constraint easier. 

\paragraph{Summary.}
The general solution to the OPE associativity constraint \eqref{OPE_Ass_LC} is
\begin{equation}
\boxed{
\begin{aligned}\label{commuting_case}
&\mathcal{C}^{1234\lambda_i}=\frac{(k^{1234}_- +(-)^{\lambda_i+\lambda_{12}}k^{1234}_+)(\Lambda-2)!}{2^{\Lambda-2}(\lambda_{12}+\lambda_i-1)!(\lambda_{34}-\lambda_i-1)!}\qquad 
    \forall\;\lambda_i\,,\quad \text{same for $(3124)$ and $(1423)$}\\ 
    & k^{1234}_-=k^{1423}_+,\qquad
    k^{3124}_-=(-)^{\Lambda}k^{1234}_+,\qquad
    k^{1423}_-=(-)^{\Lambda}k^{3124}_+\,.
\end{aligned}
}
\end{equation}
When only even-derivative vertices are present, the solution is
\begin{equation}
\boxed{
\begin{aligned}\label{commuting_even_case}
    &\mathcal{C}^{1234\lambda_i}=\frac{k^{1234}_+(\Lambda-2)!}{2^{\Lambda-3}(\lambda_{12}+\lambda_i-1)!(\lambda_{34}-\lambda_i-1)!}\qquad 
    \forall\;\lambda_i\,,\\
    &\;\text{same for $(3124)$ and $(1423)$}\qquad k^{1234}_+=k^{3124}_+=k^{1423}_+\,.
\end{aligned}
}
\end{equation}

\subsection{Colour case}
In the presence of a colour factor --- such as when fields live in representations of a gauge group $G=U(N), SO(N)$ and $USp(N)$ --- we only need to consider colour-ordered amplitudes, as discussed above. For instance, we focus here on the ordering $[1234]$, then \eqref{normal_case_2}. The associativity constraint gets modified into
\begin{align}\label{OPE_colour}
    \begin{split}
\sum_{\lambda_i}(-)^{\lambda_i}\theta_{\lambda_i}\Bigg[&\frac{\mathcal{A}_3(p_1^{\lambda_1},p_2^{\lambda_2},p_i^{\lambda_i})\mathcal{A}_3(p_3^{\lambda_3},p_4^{\lambda_4},p_i^{-\lambda_i})}{\bar{z}_{12}\bar{z}_{i3}(\omega_1+\omega_2)\omega_1\omega_2\omega_3}+\frac{\mathcal{A}_3(p_4^{\lambda_4},p_1^{\lambda_1},p_i^{\lambda_i})\mathcal{A}_3(p_2^{\lambda_2},p_3^{\lambda_3},p_i^{-\lambda_i})}{\bar{z}_{41}\bar{z}_{2i}(\omega_4+\omega_1)\omega_4\omega_1\omega_2}\Bigg]=0\,,
    \end{split}
\end{align}
where we introduce a factor $\theta_{\lambda_i}$ which can be justified by interpreting the amplitudes as arising from a Poisson bracket, with an associated $\theta_{\lambda_i}$ factor. For instance, in the $u(N)$ case, we have
\begin{equation}\label{Poisson_U(N)}
  [(\phi^{\lambda}_p)^A_{\;B},(\phi^{\mu}_q)^C_{\;D}]=\frac{\delta^{\lambda,-\mu}\delta^3(p+q)}{2p^+}\,\theta_{\lambda}\delta^C_{\;B}\delta^A_{\;D}\,,
\end{equation}
where $\theta_{\lambda}=e^{ix\lambda}$ is a phase factor reflecting a possible ambiguity in the Poisson bracket. Following the same conventions used in \cite{Serrani:2025owx,Skvortsov:2020wtf}, and in analogy with the singlet case, we fix $\theta_{\lambda_i}=(-)^{\lambda_i}$. This choice is again equivalent to taking even-helicity fields to be Hermitian and odd-helicity fields to be anti-Hermitian matrices.\footnote{In the case of Yang-Mills, we can choose to work with hermitian matrices $(T^a)^{\dag}=T^a$, then an $i$ factor would appear in the commutator $[T^a,T^b]=if^{abc}T^c$, or work with anti-hermitian matrices $(T^a)^{\dag}=-T^a$ and then $[T^a,T^b]=f^{abc}T^c$.} This choice will become particularly useful when we later search for consistent lower-spin theories satisfying the OPE associativity constraint.
Rewriting \eqref{OPE_colour} using light-cone variables, we find
\begin{equation}\label{OPE_Ass_color}
\sum_{\lambda_i}\Big(\mathcal{C}^{1234\lambda_i}\PPb_{12}^{\lambda_{12}+\lambda_i-1}\PPb_{34}^{\lambda_{34}-\lambda_i-1}-\mathcal{C}^{4123\lambda_i}\PPb_{41}^{\lambda_{14}+\lambda_i-1}\PPb_{23}^{\lambda_{23}-\lambda_i-1}\Big)=0\,.
\end{equation}
Using the variables defined in \eqref{ABC_variables}, we get 
\begin{align}\label{Ass_color}
    \begin{split}
    \sum_{\lambda_i}\Big(&\mathcal{C}^{1234\lambda_i}(A-B)^{\lambda_{12}+\lambda_i-1}(A+B)^{\lambda_{34}-\lambda_i-1}\\
    &+(-)^{\lambda_{14}+\lambda_i}\mathcal{C}^{4123\lambda_i}(C-A)^{\lambda_{14}+\lambda_i-1}(C+A)^{\lambda_{23}-\lambda_i-1}\Big)=0\,,
    \end{split}
\end{align}
and plugging in the definitions \eqref{definitions} for the functions $f^{\cdot\cdots}_-$, we have
\begin{equation}\label{Ass_color_f}
f^{1234}_-(A,B)+(-)^{\Lambda}f^{4123}_+(C,A)=0\,.
\end{equation}
As before, we can determine the form of these functions to be 
\begin{subequations}\label{eq_color}
\begin{align}\label{eq1_color}
    f^{1234}_-(A,B)&=k^{1234}_-A^{\Lambda-2}\,,\\
    f^{4123}_+(C,A)&=(-)^{\Lambda+1}k^{4123}_-A^{\Lambda-2}\,,
\end{align}
\end{subequations}
where in the definition of $k^{\cdots}_-$ we also include the $\theta_{\lambda_i}$ factor. Substituting them into \eqref{Ass_color_f} leads to the constraint 
\begin{equation}
    k_-^{1234}=k_-^{4123}\,.
\end{equation}
From \eqref{eq_color}, we find the system for the couplings:
\begin{align}
    \begin{split}
    &\mathcal{C}^{1234\lambda_i}=\frac{k_-^{1234}(\Lambda-2)!}{2^{\Lambda-2}(\lambda_{12}+\lambda_i-1)!(\lambda_{34}-\lambda_i-1)!}\quad\forall\;\lambda_i\,,\quad\text{same for $(4123)$}\,,\\
    &k^{1234}_-=\sum_{\lambda_i}\mathcal{C}^{1234\lambda_i}\,,
    \end{split}
\end{align}
Let us conclude with a comment on the possibility of using $SU(N)$ as the gauge group. In this case, the Poisson bracket in \eqref{Poisson_U(N)} acquires an additional term of the form $\sim \frac{1}{N}\delta^A_{\;B}\delta^C_{\;D}$. This contribution induces, in the holomorphic constraint, a term of the type \eqref{mixed_case_2}. As a result, the solution to the constraint becomes more involved. However, in special cases, such as for the cubic vertices of Yang–Mills theory, where the couplings are proportional to the fully antisymmetric structure constants $f_{abc}$ this additional term vanishes. Indeed, one would obtain an extra contribution of the form $\sim \mathrm{Tr}([T_{a_1},T_{a_2}])\mathrm{Tr}([T_{a_3},T_{a_4}])(\phi^{\lambda_1}_{q_1})^{a_1}(\phi^{\lambda_2}_{q_2})^{a_2}(\phi^{\lambda_3}_{q_3})^{a_3}(\phi^{\lambda_4}_{q_4})^{a_4}$, which vanishes due to the cyclicity of the trace. This mechanism is sometimes referred to as photon decoupling: at tree level, gluon scattering amplitudes are identical whether one works with $U(N)$ or $SU(N)$.

\paragraph{Summary.} The solution to the associativity constraint in the colour-ordered case is
\begin{equation}\label{colour_case}
    \boxed{
    \begin{aligned}
    &\mathcal{C}^{1234\lambda_i}=\frac{k_-^{1234}(\Lambda-2)!}{2^{\Lambda-2}(\lambda_{12}+\lambda_i-1)!(\lambda_{34}-\lambda_i-1)!}\qquad\forall\;\lambda_i\,,\\
    &\;\text{same for $(4123)$}
    \qquad k^{1234}_-=k^{4123}_-\,.
    \end{aligned}
    }
\end{equation}

\subsection{Main mechanism}
As we have seen above, there is a simple procedure to solve the constraints. By looking at the most general case presented in Appendix \ref{Paper2_AppendixD}, one can see that this is always the case.
We can divide the procedure into two parts. First, we see that the constraint reduces to setting a polynomial in three variables to zero. The admissible functions available for this purpose are, as indicated in \eqref{possible_functions}, the following:
\begin{align}\label{fAB}
    f(A,B)&= a_1 A^{\Lambda-2}+ b_1 B^{\Lambda-2}\,,& \mathcal{C}^{1234\lambda_i},\,\mathcal{C}^{2134\lambda_i},\,\mathcal{C}^{1243\lambda_i},\,\mathcal{C}^{2143\lambda_i}\,,\\\label{fBC}
    f(B,C)&= b_2 B^{\Lambda-2}+ c_1 C^{\Lambda-2}\,,& \mathcal{C}^{1342\lambda_i},\,\mathcal{C}^{3142\lambda_i},\,\mathcal{C}^{1324\lambda_i},\,C^{3124\lambda_i}\,,\\\label{fCA}
    f(C,A)&= c_2 C^{\Lambda-2}+ a_2 A^{\Lambda-2}\,,& \mathcal{C}^{1423\lambda_i},\,\mathcal{C}^{4123\lambda_i},\,\mathcal{C}^{1432\lambda_i},\,\mathcal{C}^{4132\lambda_i}\,,
\end{align}
where on the right of each function, we wrote the product of couplings that can produce them. The constraint they have to satisfy is 
\begin{equation}\label{cocycle?_}
    f(A,B)+f(B,C)+f(C,A)=0\,.
\end{equation}
These polynomials can cancel each other in different ways to ensure the total sum vanishes:
\begin{enumerate}
    \item If all coefficients are non-zero, consistency requires setting $a_1=-a_2$, $b_1=-b_2$, and $c_1=-c_2$. This corresponds to the case of singlet fields, where only even-derivative vertices are present, as shown in \eqref{commuting_even_case}. In this case, we need at least one product of couplings in \eqref{fAB}, in \eqref{fBC}, and in \eqref{fCA} to be present.

    \item If only two terms are non-zero --- for instance, $b_1=b_2=c_1=c_2=0$ and $a_1=-a_2$ --- we obtain a solution that corresponds to the colour case \eqref{colour_case}; the same holds under permutations of $a,b,c$. In this case, we need at least one product of couplings in \eqref{fAB} and \eqref{fCA} to be present.

    \item One can consider configurations with two vanishing terms and four non-zero coefficients, for example, $c_1=c_2=0$, $a_1=-a_2$, and $b_1=-b_2$. This type of cancellation may appear in the singlet case as well, as in \eqref{commuting_case}, but it requires the inclusion of odd-derivative vertices. In this case, we need at least one product of couplings in \eqref{fAB}, \eqref{fBC}, and \eqref{fCA} to be present.
\end{enumerate}
Once one of these cases is realised, by looking at the form of a single function, for instance, $f(A,B)= a_1 A^{\Lambda-2}+ b_1 B^{\Lambda-2}$, and recalling the definitions \eqref{definitions}, we can extract the solutions for the couplings and find two distinct cases:
\begin{enumerate}
    \item If we have $a_1\neq 0$ and $b_1\neq 0$, to reproduce the correct form of the function, we must sum over all even or odd integer values of $\lambda_i$.

    \item If we have $a_1\neq 0$ and $b_1=0$, or vice versa, to reproduce the correct form of the function, we must sum over all integer values of $\lambda_i$.
\end{enumerate}

\subsection{Mixed cases}
Here we analyse the three remaining cases \eqref{mixed_case_1}, \eqref{mixed_case_2}, and \eqref{mixed_case_3}. We start by looking at \eqref{mixed_case_1} then
\begin{equation}
    \mathrm{Tr}(T_{a_1}T_{a_2})(\phi^{\lambda_1}_{q_1})^{a_1}(\phi^{\lambda_2}_{q_2})^{a_2}\phi^{\lambda_3}_{q_3}\phi^{\lambda_4}_{q_4}\sim \delta_{a_1a_2}(\phi^{\lambda_1}_{q_1})^{a_1}(\phi^{\lambda_2}_{q_2})^{a_2}\phi^{\lambda_3}_{q_3}\phi^{\lambda_4}_{q_4}\,.
\end{equation}
In this case, all fields ``commute'', as in the case of singlet fields. Therefore, the solution to the constraint coincides with the one found in the singlet case \eqref{commuting_case} and \eqref{commuting_even_case}.
Regarding \eqref{mixed_case_2} we have
\begin{equation}
    \mathrm{Tr}(T_{a_1}T_{a_2})\mathrm{Tr}(T_{a_3}T_{a_4})(\phi^{\lambda_1}_{q_1})^{a_1}(\phi^{\lambda_2}_{q_2})^{a_2}(\phi^{\lambda_3}_{q_3})^{a_3}(\phi^{\lambda_4}_{q_4})^{a_4}\sim \delta_{a_1a_2}\delta_{a_3a_4}(\phi^{\lambda_1}_{q_1})^{a_1}(\phi^{\lambda_2}_{q_2})^{a_2}(\phi^{\lambda_3}_{q_3})^{a_3}(\phi^{\lambda_4}_{q_4})^{a_4}
\end{equation}
In this case, we have to consider the following four exchanges
\begin{align}
    &(1234)+(2134)+(1243)+(2143)\,,
\end{align}
where for each $(1234)$ ordering we also include the associated $(3412)$ exchange. As shown above, in the presence of these exchanges, the holomorphic quartic constraint, reducing to the equation \eqref{cocycle?_}, forces $f(A,B)=0$, which cannot be satisfied. This conclusion holds for generic vertices, and in the absence of additional vertices that would contribute extra terms to \eqref{cocycle?_}. We now consider the last case, \eqref{mixed_case_3}. We have
\begin{equation}
    \mathrm{Tr}(T_{a_1}T_{a_2}T_{a_3})(\phi^{\lambda_1}_{q_1})^{a_1}(\phi^{\lambda_2}_{q_2})^{a_2}(\phi^{\lambda_3}_{q_3})^{a_3}\phi^{\lambda_4}_{q_4}\,.
\end{equation}
In this case, using the cyclic property of the trace, and that $\phi^{\lambda_4}_{q_4}$ ``commutes'' with the others, we have to consider the following exchanges
\begin{align}
    &(123)4+(312)4+(231)4&\rightarrow&
    &(1234)+(1243)+(3142)+(3124)+(1423)+(4123)\,,
\end{align}
where again for each $(1234)$ we also have the associated $(3412)$ exchange. To read off the solution, we can look at the general case in Appendix \ref{Paper2_AppendixD}, and we get 
\begin{subequations}\label{mixed_constraint}
\begin{align}
    &(\mathcal{F}^{1234}+\mathcal{F}^{1243})k^{1234}_- - (\mathcal{F}^{1423}+\mathcal{F}^{4123})k^{1423}_+=0\;\implies\;\mathcal{F}^{1234}k^{1234}_- = \mathcal{F}^{1423}k^{1423}_+\,,\\
    &(\mathcal{F}^{1342}+\mathcal{F}^{1342})k^{1342}_- - (\mathcal{F}^{1234}+\mathcal{F}^{1243})k^{1234}_+=0\;\implies\;\mathcal{F}^{1342}k^{1342}_- = \mathcal{F}^{1234}k^{1234}_+\,,\\
    &(\mathcal{F}^{1423}+\mathcal{F}^{4123})k^{1423}_- - (\mathcal{F}^{1342}+\mathcal{F}^{1342})k^{1342}_+=0\;\implies\;\mathcal{F}^{1423}k^{1423}_- = \mathcal{F}^{1342}k^{1342}_+\,,
\end{align}
\end{subequations}
where in the second expression we identified the various $\mathcal{F}$ terms since they are equal in this case, like $\mathcal{F}^{1234}=f_{a_1a_2a_3}=\mathcal{F}^{1243}$, and the value of the coupling product $\mathcal{C}^{1234}$, as well as for $(1342)$ and $(1423)$ is given in \eqref{usual_couplings}.
If we consider only even-derivative vertices (i.e. with $k^{\cdots\cdot}_-=k^{\cdots\cdot}_+$), we get
\begin{equation}\label{mixed_only_even}
   \mathcal{F}^{1234}k^{1234}_-=\mathcal{F}^{1342}k^{1342}_-=\mathcal{F}^{1423}k^{1423}_-=\mathcal{F}^{1234}k^{1234}_+=\mathcal{F}^{1342}k^{1342}_+=\mathcal{F}^{1423}k^{1423}_+\,.
\end{equation}
Another particular solution to the constraint in \eqref{mixed_constraint} arises when we start with a pair of couplings $C^{\lambda_1,\lambda_2,\lambda_i}C^{-\lambda_i,\lambda_3,\lambda_4}$  where one is an even-derivative and the other an odd-derivative vertex. Borrowing ideas from \cite{Serrani:2025owx}, we can rewrite the solution in a more convenient form using
\begin{align}\label{new_variables}
    k^{1234}_E\definition \frac{1}{2}(k^{1234}_-+k^{1234}_+)&\equiv\sum_{(\lambda_{12}+\lambda_i)\in\,\text{even}}\mathcal{C}^{1234\lambda_i}\,,\\
    k^{1234}_O\definition\frac{1}{2}(k^{1234}_--k^{1234}_+)&\equiv \sum_{(\lambda_{12}+\lambda_i)\in\,\text{odd}}\mathcal{C}^{1234\lambda_i}\,.
\end{align}
We can then rewrite the system governing the couplings in the form
\begin{align}\label{even_first_P2}
    \mathcal{C}^{1234\lambda_i}&=\frac{k^{1234}_E (\Lambda-2)!}{2^{\Lambda-3}(\lambda_{12}+\lambda_i-1)!(\lambda_{34}-\lambda_i-1)!}\qquad 
    \forall\;(\lambda_{12}+\lambda_i)\in\text{even}\,,\\ \label{odd_first_P2}
    \mathcal{C}^{1234\lambda_i}&=\frac{k^{1234}_O (\Lambda-2)!}{2^{\Lambda-3}(\lambda_{12}+\lambda_i-1)!(\lambda_{34}-\lambda_i-1)!}\qquad 
    \forall\;(\lambda_{12}+\lambda_i)\in\text{odd}\,.
\end{align}
If we rewrite the constraint \eqref{mixed_constraint} in terms of these new variables, we find 
\begin{align}\label{mixed_odd_even}
    \nonumber
   &\mathcal{F}^{1234}(k^{1234}_E+k^{1234}_O) = \mathcal{F}^{1423}(k^{1423}_E-k^{1423}_O)\,,\quad
    \mathcal{F}^{1342}(k^{1342}_E+k^{1342}_O) = \mathcal{F}^{1234}(k^{1234}_E-k^{1234}_O)\,,\\
    &\mathcal{F}^{1423}(k^{1423}_E+k^{1423}_O) = \mathcal{F}^{1342}(k^{1342}_E-k^{1342}_O)\,,
\end{align}
and becomes easier to identify the two special solutions. The first occurs when $k^{\cdots}_O=0$, leaving only $k^{\cdots}_E$, which corresponds to the purely even-derivative case discussed in \eqref{mixed_only_even}, and gives
\begin{equation}\label{mixed_1}
\mathcal{F}^{1234}k^{1234}_E=\mathcal{F}^{1342}k^{1342}_E=\mathcal{F}^{1423}k^{1423}_E\,.
\end{equation}
The second special case arises when two of the $k^{\cdots}_E$ coefficients and one $k^{\cdots}_O$ term vanish. For instance, setting $k^{1234}_E=k^{1342}_E=k^{1423}_O=0$ we find
\begin{equation}\label{mixed_2}
    \mathcal{F}^{1234}k^{1234}_O =-\mathcal{F}^{1342}k^{1342}_O=\mathcal{F}^{1423}k^{1423}_E\,.
\end{equation}
Notice that the same solution also applies to the constraint found in the singlet case \eqref{commuting_case}, provided we drop the $\mathcal{F}$ terms. On the other hand, if we look for solutions involving only odd-derivative vertices (i.e. with $k^{\cdots\cdot}_-=-k^{\cdots\cdot}_+$), we encounter a contradiction. This is evident from the structure of \eqref{mixed_constraint}, and was already shown for the singlet case in \cite{Serrani:2025owx}.

\section{Solutions to the OPE associativity}\label{Paper2_section5}

In the section above, we solved the OPE associativity constraint in several specific cases, with the most general case addressed in Appendix \ref{Paper2_AppendixD}. However, finding a consistent theory composed of holomorphic cubic couplings that simultaneously solves all possible OPE associativity constraints arising among them is a related but distinct problem.

Fortunately, this issue was already addressed in \cite{Serrani:2025owx} for the holomorphic light-cone constraint, and the same approach applies here. The key idea is that, starting from a specific pair of cubic couplings $C^{\lambda_1,\lambda_2,\lambda_i}C^{-\lambda_i,\lambda_3,\lambda_4}$ involving at least one opposite helicity field,\footnote{Note that the OPE associativity constraint becomes non-trivial only if at least two of the couplings involved contain fields of opposite helicity, i.e. $\lambda_i = -\lambda_j$. In principle, one could consider very generic couplings that never generate any non-trivial constraint, though such theories would be rather trivial. Indeed, recall that to obtain an exchange amplitude, we need at least two cubic couplings involving opposite helicity fields.} requiring the constraint to be satisfied may --- and typically does --- generate new cubic couplings. These, in turn, can give rise to further constraints, and the process continues recursively. For a detailed discussion, see \cite{Serrani:2025owx}.

We briefly explain here the procedure to find consistent solutions for various cases. In the case of a singlet field \eqref{commuting_even_case} and the mixed case \eqref{mixed_1} with only even-derivative vertices, satisfying the OPE associativity requires that each time one of the pairs of couplings below is present, all the others must follow:
\begin{align}\label{singlet_theory}
    \begin{split}
    &C^{\lambda_1,\lambda_2,[2-\lambda_{12},k-\lambda_{12}]}C^{[\lambda_{12}-2,\lambda_{12}-k],\lambda_3,\lambda_4}\,,\qquad
    C^{\lambda_1,\lambda_3,[2-\lambda_{13},k-\lambda_{13}]}C^{[\lambda_{13}-2,\lambda_{13}-k],\lambda_2,\lambda_4}\,,\\
    &C^{\lambda_1,\lambda_4,[2-\lambda_{14},k-\lambda_{14}]}C^{[\lambda_{14}-2,\lambda_{14}-k],\lambda_2,\lambda_3}\,,
    \end{split}
\end{align}
where we write them in pairs to emphasize that each involves an opposite-helicity field, $k$ is an even integer, the square bracket $[-,-]$ is a notation to indicate the range of exchanged helicities, incremented by steps of $2$, and $\Lambda=\lambda_1+\lambda_2+\lambda_3+\lambda_4$ --- the total number of derivatives --- must be even. For a better explanation of the notation used, we refer to \cite{Serrani:2025owx}. For instance, the following 
\begin{align}
    \begin{split}
    &C^{\lambda_1,\lambda_2,[2-\lambda_{12},4-\lambda_{12}]}C^{[\lambda_{12}-2,\lambda_{12}-4],\lambda_3,\lambda_4}\,,\qquad
    C^{\lambda_1,\lambda_3,[2-\lambda_{13},4-\lambda_{13}]}C^{[\lambda_{13}-2,\lambda_{13}-4],\lambda_2,\lambda_4}\,,\\
    &C^{\lambda_1,\lambda_4,[2-\lambda_{14},4-\lambda_{14}]}C^{[\lambda_{14}-2,\lambda_{14}-4],\lambda_2,\lambda_3}\,,
    \end{split}
\end{align}
would correspond to the following $6$ pairs of coupling
\begin{align}
    \begin{split}
    &C^{\lambda_1,\lambda_2,2-\lambda_{12}}C^{\lambda_{12}-2,\lambda_3,\lambda_4}\,,\qquad
    C^{\lambda_1,\lambda_3,2-\lambda_{13}}C^{\lambda_{13}-2,\lambda_2,\lambda_4}\,,\qquad
    C^{\lambda_1,\lambda_4,2-\lambda_{14}}C^{\lambda_{14}-2,\lambda_2,\lambda_3}\,,\\
    &C^{\lambda_1,\lambda_2,4-\lambda_{12}}C^{\lambda_{12}-4,\lambda_3,\lambda_4}\,,\qquad
    C^{\lambda_1,\lambda_3,4-\lambda_{13}}C^{\lambda_{13}-4,\lambda_2,\lambda_4}\,,\qquad
    C^{\lambda_1,\lambda_4,4-\lambda_{14}}C^{\lambda_{14}-4,\lambda_2,\lambda_3}\,.
    \end{split}
\end{align}
For the colour case \eqref{colour_case} with both even- and odd-derivative vertices, we need the following set of couplings:
\begin{align}\label{colour_theory}
    \begin{split}
    &C^{\lambda_1,\lambda_2,[1-\lambda_{12},k-\lambda_{12}]}C^{[\lambda_{12}-1,\lambda_{12}-k],\lambda_3,\lambda_4}\,,\qquad
    C^{\lambda_1,\lambda_3,[1-\lambda_{13},k-\lambda_{13}]}C^{[\lambda_{13}-1,\lambda_{13}-k],\lambda_2,\lambda_4}\,,\\
    &C^{\lambda_1,\lambda_4,[1-\lambda_{14},k-\lambda_{14}]}C^{[\lambda_{14}-1,\lambda_{14}-k],\lambda_2,\lambda_3}\,,
    \end{split}
\end{align}
where $k$ is an integer, and the square bracket $[-,-]$ is incremented by steps of $1$. For the mixed case \eqref{mixed_2}, we need the following set of couplings:
\begin{align}\label{mixed2_theory}
    \begin{split}
    &C^{\lambda_1,\lambda_2,[1-\lambda_{12},k-\lambda_{12}]}C^{[\lambda_{12}-1,\lambda_{12}-k],\lambda_3,\lambda_4}\,,\qquad
    C^{\lambda_1,\lambda_3,[1-\lambda_{13},k-\lambda_{13}]}C^{[\lambda_{13}-1,\lambda_{13}-k],\lambda_2,\lambda_4}\,,\\
    &C^{\lambda_1,\lambda_4,[k-\lambda_{14},1-\lambda_{14}]}C^{[\lambda_{14}-k,\lambda_{14}-1],\lambda_2,\lambda_3}\,,
    \end{split}
\end{align}
where $k$ is an even integer, the square bracket $[-,-]$ is incremented by steps of $2$, and $\Lambda$ must be odd.

Once all the couplings required to satisfy the constraint are included, an explicit solution relating the various couplings can always be extracted from \eqref{commuting_even_case}, \eqref{colour_case}, \eqref{even_first_P2}, and \eqref{odd_first_P2}. Below, we write down some explicit solutions for the couplings. In particular, we match the result of \cite{Ren:2022sws} for lower-spin theories and pinpoint the difference with the solutions to the light-cone quartic holomorphic constraint. 

\subsection{Lower-spin solutions}
Following ideas from \cite{Metsaev:1991mt,Metsaev:1991nb,Ponomarev:2016lrm,Ponomarev:2017nrr,Serrani:2025owx} we look for the most general lower-spin cubic vertices (i.e. involving lower-spins $|\lambda|=0,1,2$) that solve the OPE associativity constraint.

\paragraph{Singlet fields.} This presentation closely follows that of \cite{Serrani:2025owx}, with the necessary modifications. All possible even-derivative couplings are given by
\begin{equation}
    \left\{C^{-2,2,2},C^{-1,1,2},C^{0,0,2},C^{0,1,1},C^{0,2,2},C^{1,1,2},C^{2,2,2}\right\}\,.
\end{equation}
In total, we have $7$ couplings: $5$ abelian and $2$ non-abelian.\footnote{The non-abelian vertices are the most interesting ones, as they deform the gauge algebra in a covariant formulation. The abelian ones do not. Nevertheless, it is worth noting that, in general, abelian vertices can also be constrained and may play an important role in ensuring the consistency of the theory. Since we work in the light-cone gauge and with holomorphic theories, we define abelian couplings as those with $\lambda_i\geq0$ and $\sum_i\lambda_i>0$. Note that $(0,s,s)$ is still considered abelian this way, which is justified since the spin-zero exchange disappears from the constraint. Such couplings are, obviously, consistent on their own unless there are non-abelian couplings that ``talk to them''. } We can now search for all possible solutions to the OPE associativity, following \eqref{singlet_theory}.

We begin with two-derivative couplings, then $\left\{C^{-2,2,2},C^{-1,1,2},C^{0,0,2},C^{0,1,1}\right\}$.
Solutions to the OPE associativity are
\begin{subequations}
\begin{align}
        &\{C^{-2,2,2}\}\,,&&\text{graviton coupling}\,,\\
        &\{C^{1,1,0}\}\,,&&\text{photons coupled to scalars via } \phi F_{\mu\nu}^2\,,\\
        &\{C^{-2,2,2}=C^{0,0,2}\}\,,&&\text{scalars coupled to graviton}\,,\\
        &\{C^{-2,2,2}=C^{-1,1,2}\}\,,&&\text{graviton coupled to photons}\,,\\
        &\{C^{-2,2,2}=C^{-1,1,2}=C^{0,0,2},C^{0,1,1}\}\,,&& \text{graviton, photons and scalars}\,.
\end{align}
\end{subequations}
We use the following notation: within $\{\cdots\}$, we list the active (free) couplings. When a coupling appears alone, it is considered unconstrained --- that is, it can appear in the theory’s action with a free coefficient. It is important to specify the active couplings before solving the constraints. Note that the relations above are consistent with the universality of gravitational interactions.

If we allow for higher-derivative terms, the most general solution we obtain is the following:
\begin{equation}\label{HDlowerspin_P2}
    \left\{C^{-2,2,2}=C^{-1,1,2}=C^{0,0,2},C^{1, 1, 2}=\frac{C^{0, 1, 1} C^{0, 2, 2}}{C^{-2, 2, 2}},C^{2, 2, 2}=\frac{3}{10}\frac{(C^{0, 2, 2})^2}{C^{-2, 2, 2}}\right\}\,.
\end{equation}
The constraint above implies that a theory containing  $R^3$ and $RF^2$ terms with arbitrary coefficients would not admit a celestial dual, due to the failure of OPE associativity. Instead, the coefficients must be fixed to the specific values given above. These results correctly reproduce those of \cite{Ren:2022sws}. The constraint imposed to $C^{0, 1, 1}$ in \eqref{HDlowerspin_P2} for the higher-derivative theories
\begin{equation}
    C^{1, 1, 2}=\frac{C^{0, 1, 1} C^{0, 2, 2}}{C^{-2, 2, 2}}\,,
\end{equation}
is novel, because in \cite{Ren:2022sws} the $C^{0, 1, 1}$ abelian coupling between photons and scalars was not considered.

\paragraph{Colour case.} This presentation closely follows that of \cite{Serrani:2025owx}, with the necessary modifications. We now examine all solutions in the presence of a gauge group, following \eqref{colour_theory}. The complete set of even- and odd-derivative couplings that can be constructed is given by
\begin{align}
\begin{split}
    \{&C^{-2,1,2},\textcolor{blue}{C^{-2,2,2}},\textcolor{blue}{C^{-1,0,2}},C^{-1,1,1},\textcolor{blue}{C^{-1,1,2}},\textcolor{red}{C^{-1,2,2}},C^{0,0,1},\textcolor{blue}{C^{0,0,2}},\\
&C^{0,1,1},\textcolor{blue}{C^{0,1,2}},\textcolor{blue}{C^{0,2,2}},C^{1,1,1},\textcolor{red}{C^{1,1,2}},\textcolor{red}{C^{1,2,2}},\textcolor{red}{C^{2,2,2}}\}\,.
\end{split}
\end{align}
In total, we have $15$ couplings: $9$ abelian and $6$ non-abelian. We highlight the lower-spin couplings as follows: blue indicates those which, on their own, require higher-spin couplings for consistency; red denotes those which, when combined with any other coupling, also require higher-spin ones; uncoloured couplings are, as we will show in more detail below, the only consistent lower-spin couplings. We now search for possible theories that satisfy the OPE associativity constraint. We begin by considering one-derivative interactions, involving the subset of couplings
\begin{equation}\label{one_derivative_colour}
    \{C^{-2,1,2},C^{-1,0,2},C^{-1,1,1},C^{0,0,1}\}\,.
\end{equation}
Theories that satisfy the OPE associativity constraint are given by
\begin{subequations}
\begin{align}
    &\{C^{-1,1,1}\}\,, && \text{Yang-Mills coupling}\,,\\
    &\{C^{-1,1,1}=C^{0,0,1}\}\,,&& \text{Yang-Mills coupled to scalars}\,,\\\label{coloured_graviton}
    &\{C^{-2,1,2}=C^{-1,1,1}\}\,,&& \text{coloured graviton}\,,
\end{align}
\end{subequations}
where we do not have any sign factor appearing, thanks to the use of the $\theta_{\lambda_i}=(-)^{\lambda_i}$ term in \eqref{OPE_colour}. This also matches standard definitions for lower-spin theories and with the results in \cite{Ren:2022sws}.
Note that including $C^{-1,0,2}$ and at least one other of the couplings in \eqref{one_derivative_colour} leads to a non-associative OPE.

Another interesting option is the coloured graviton. It is well known that, under the usual assumptions, gravitons cannot carry colour \cite{Boulanger:2000rq}. However, we find that it is possible to have multi-graviton theories that satisfy the OPE associativity, provided we restrict to the self-dual case. The coupling $C^{-2,1,2}$ is the ``usual'' current interaction of one helicity of the Yang-Mills field with a current built from a multiplet of gravitons.

Including higher-derivative couplings, the most general consistent solution is
\begin{equation}\label{HDcolortheory}
\Big\{C^{-1,1,1}=C^{-2,1,2}=C^{0,0,1},C^{1,1,1}=\frac{1}{2}\frac{(C^{0, 1, 1})^2}{C^{-1, 1, 1}}\Big\}\,.
\end{equation}
In this case as well, the associativity constraint implies that a theory containing an $F^3$ term with an arbitrary coefficient would not admit a celestial dual, due to the failure of OPE associativity. Any attempt to include other couplings breaks OPE associativity, forcing us to introduce higher-spin fields. This is consistent with expectations from the covariant formalism. For example, the coupling $C^{-2,2,2}$ would lead to ``colour gravity'', which is known not to be consistent, at least when including only lower spins. These results correctly reproduce those of \cite{Ren:2022sws}, once we use notation valid for any semisimple Lie algebra given in \eqref{semisimple_a}--\eqref{semisimple_b}. 
The possibility of having OPE associativity in the presence of the non-abelian $C^{-2, 1, 2}$ coupling \eqref{coloured_graviton} is novel because in \cite{Ren:2022sws} the exotic $C^{-2, 1, 2}$ non-abelian coupling between gravitons and spin-$1$ fields was not included.

\paragraph{Other solutions for lower-spin.} Some vertices are still missing to fully reproduce the lower-spin results obtained in \cite{Ren:2022sws}. In particular, we need to study the OPE associativity constraints imposed on the following couplings, which were not included in the analysis above
\begin{equation}
    \left\{\delta_{ab}\frac{\tilde{C}^{0,1,1}\PPb^2}{\beta_2\beta_3}\phi^0_{q_1}(\phi^1_{q_2})^a(\phi^1_{q_3})^b,\delta_{ab}\frac{\tilde{C}^{1,1,2}\PPb^4}{\beta_1^2\beta_2\beta_3}\phi^2_{q_1}(\phi^1_{q_2})^a(\phi^1_{q_3})^b,\delta_{ab}\frac{\tilde{C}^{-1,1,2}\PPb^2}{\beta_1^2\beta_2^{-1}\beta_3}\phi^2_{q_1}(\phi^{-1}_{q_2})^a(\phi^1_{q_3})^b\right\}\,,
\end{equation}
where we denote the coupling by $\tilde{C}$ to distinguish it from the other $C$'s.
To treat these cases and match the result in \cite{Ren:2022sws}, we need to use the solutions found for the mixed cases. For the mixed cases \eqref{mixed_case_1}, following \eqref{singlet_theory} and using \eqref{commuting_even_case} we can find
\begin{align}
    &(C^{-2,2,2}-\tilde{C}^{-1,1,2})\tilde{C}^{-1,1,2}=0\,,&
    &(C^{-2,2,2}-\tilde{C}^{-1,1,2})\tilde{C}^{1,1,2}=0\,,\\
    &(C^{0,0,2}-\tilde{C}^{-1,1,2})\tilde{C}^{0,1,1}=0\,,&
    &\tilde{C}^{-1,1,2}\tilde{C}^{1,1,2}=C^{0,2,2}\tilde{C}^{0,1,1}\,.
\end{align}
While for the mixed cases \eqref{mixed_case_3}, following \eqref{mixed2_theory} and using \eqref{mixed_2} we can find
\begin{align}
\begin{split}
    \frac{1}{2!}f_{a_1a_2c}C^{1,1,-1}\delta_{c\,a_3}\tilde{C}^{1,1,2}=-\frac{1}{2!}f_{a_1a_3c}C^{1,1,-1}\delta_{c\,a_2}\tilde{C}^{1,1,2}=\frac{1}{2!}\delta_{a_1c}\tilde{C}^{1,2,1}f_{c\,a_2a_3}C^{-1,1,1}\\
    =\frac{1}{3!}f_{a_1a_2c}C^{1,1,1}\delta_{c\,a_3}\tilde{C}^{-1,1,2}=-\frac{1}{3!}f_{a_1a_3c}C^{1,1,1}\delta_{c\,a_2}\tilde{C}^{-1,1,2}=\frac{1}{3!}\delta_{a_1c}\tilde{C}^{1,2,-1}f_{c\,a_2a_3}C^{1,1,1}\,.
    \end{split}
\end{align}
Then, using the full antisymmetry of $f_{abc}$ and the fact that the spin-$2$ field is a singlet, we get
\begin{equation}
    f_{a_1a_2a_3}\tilde{C}^{-1,1,2}C^{1,1,1}=3f_{a_1a_2a_3}\tilde{C}^{1,1,2}C^{-1,1,1}\implies\tilde{C}^{-1,1,2}C^{1,1,1}=3\,\tilde{C}^{1,1,2}C^{-1,1,1}\,.
\end{equation}
In principle, there may be --- and likely are --- other admissible lower-spin theories that satisfy OPE associativity, but we do not analyse them further here.

We conclude this section by noting that this method of solving the OPE associativity constraints substantially simplifies the analysis compared to \cite{Mago:2021wje,Ren:2022sws}. Furthermore, it enables extensions to higher-spin fields and to solutions with higher-derivative cubic couplings, as discussed in \cite{Serrani:2025owx}.

\section{Holomorphic constraint vs OPE associativity}\label{Paper2_section6}

In this section, we make the connection between the quartic light-cone holomorphic constraint and the holomorphic OPE associativity more explicit. In particular, we borrow the idea of a \textit{``gauge'' algebra} \cite{Ponomarev:2017nrr} constructed starting from the cubic chiral vertices. First, we review and define this Lie algebra. Then we relate the following four objects: the Jacobi identity for the gauge algebra, the four-point tree-level amplitude constructed out of the cubic chiral vertices, the OPE associativity in CCFT, and the quartic light-cone holomorphic constraint.

\subsection{A gauge algebra from cubic vertices}
In \cite{Ponomarev:2017nrr}, the author proposed a gauge algebra built from the holomorphic vertices (including the chiral higher-spin theories). This idea was originally introduced in the context of self-dual Yang-Mills (SDYM) and self-dual gravity (SDGR) in \cite{Monteiro:2011pc}, to derive the BCJ relations for these theories. In particular, they identified the relevant gauge algebra of SDGR as the Poisson algebra of area-preserving diffeomorphisms of $S^2$, and related it to the kinematical algebra of SDYM.

Let us consider the most general cubic couplings, but omit $\fA_{abc}$ whenever it is not crucial. Starting from a chiral vertex in \eqref{cubic_hamiltonian} (specifically, the holomorphic one), we associate to it a structure constant as follows:
\begin{align}
    &h_{A_1A_2A_3}=-C^{\lambda_1,\lambda_2,\lambda_3}\frac{\PPb_{12}^{\lambda_{123}}}{\beta_1^{\lambda_1}\beta_2^{\lambda_2}\beta_3^{\lambda_3}}\delta\Big(\sum_i q_i\Big)&
    &\rightarrow&
    &\fB_{A_1A_2A_3}\definition \frac{1}{2}h_{A_1A_2A_3}\frac{\beta_2\beta_3}{\beta_1\PPb_{12}}\,,
\end{align}
where capital Latin letters such as $A_i$ collectively denote helicity, momentum, and possible other internal labels. Let us note that this procedure inverts the parity of the original coupling. Effectively, it does send a cubic vertex $H_3^{\lambda_1,\lambda_2,\lambda_3}$ to a shifted one $H_3^{\lambda_1+1,\lambda_2-1,\lambda_3-1}$, thereby reversing the parity of the coupling. This way, the structure constants read
\begin{equation}\label{structure_const}
\fB_{A_1A_2A_3}=-\frac{1}{2}C^{\lambda_1,\lambda_2,\lambda_3}\frac{\PPb_{12}^{\lambda_{123}-1}}{\beta_1^{\lambda_1+1}\beta_2^{\lambda_2-1}\beta_3^{\lambda_3-1}}\delta\Big(\sum_i q_i\Big)\,. 
\end{equation}
We raise and lower indices using the natural inner product
\begin{equation}
    (\phi_1,\phi_2)\definition \sum_{\lambda,s}\int d^4q\, d^4p\,\delta^4(q+p)\,\delta_{ab}\,\delta^{\lambda,-s}(\phi^{\lambda}_q)^a(\phi^{s}_p)^b\,,
\end{equation}
where $\phi$ are the fields of the theory. Notice that, according to the definition of the inner product, raising and lowering indices correspond to flipping the sign of both helicity and momentum. Instead, nothing happens to internal indices, if present, due to the metric normalized to $\delta_{ab}$. For any chiral theory, the cubic chiral action can be written as
\begin{equation}
    S=\frac{1}{2}(\phi,\Box\phi)-h_{A_1A_2A_3}\phi^{A_1}\phi^{A_2}\phi^{A_3}\,.
\end{equation}
The idea to define $\fB_{ABC}$ is to factor out the kinematical part of the self-dual Yang-Mills vertex, and interpret other theories, e.g. the Chiral higher-spin theories, as generalisations of SDYM \cite{Ponomarev:2017nrr} with a certain gauge algebra defined by $\fB_{ABC}$. If we introduce the Lie bracket $[\bullet,\bullet]$ 
\begin{equation}
    [T_{A_1},T_{A_2}]=\fB^{A_3}_{\;\;\;\;A_1A_2}T_{A_3}\,,
\end{equation}
that corresponds to $\fB_{ABC}$, we can rewrite the action as the ``SDYM'' where the colour Lie algebra is replaced by the gauge algebra.\footnote{We write ``SDYM'' since theories can have fields of various helicities. It is also important that there is no scalar self-coupling for such an interpretation to work (this is the only coupling from which $\PPb$ cannot be factored out). One can also consider SDYM/SDGR with these gauge algebras. For example, the Moyal deformation of SDGR was discussed recently in \cite{Bu:2022iak}. Such theories would break Lorentz invariance since the number of $\PPb$ in a vertex is strongly correlated with the helicities. }   

It is interesting to see whether $\fB_{A_1A_2A_3}$ defines a Lie algebra. We need to check the Jacobi identity, which gives a constraint on the couplings. As a warm-up example, let us review what we get for SDYM \cite{Ponomarev:2017nrr}. We could start from the Chalmers-Siegel action \cite{Chalmers:1996rq} as in \cite{Ponomarev:2017nrr} and assume to activate the SDYM coupling $C^{-1,1,1}$.
Using the following SDYM vertex:
\begin{equation}
    h_{A_1A_2A_3}=2f_{a_1a_2a_3}\frac{\PPb_{12}\beta_1}{\beta_2\beta_3}\delta_{\lambda_1,-1}\delta_{\lambda_2,1}\delta_{\lambda_3,1}\delta\Big(\sum_i q_i\Big)\,.
\end{equation}
From \eqref{structure_const} we can read the structure constant for the gauge algebra
\begin{equation}
\fB_{A_1A_2A_3}=f_{a_1a_2a_3}\big(\delta_{\lambda_1,-1}\delta_{\lambda_2,1}\delta_{\lambda_3,1}+\frac{\beta_2^2}{\beta_1^2}\delta_{\lambda_1,1}\delta_{\lambda_2,-1}\delta_{\lambda_3,1}+\frac{\beta_3^2}{\beta_1^2}\delta_{\lambda_1,1}\delta_{\lambda_2,1}\delta_{\lambda_3,-1}\big)\delta\Big(\sum_i q_i\Big)\,,
\end{equation}
and the Jacobi identity is a natural consequence of the one for the structure constant $f^{a_1a_2a_3}$.

We can do the same for SDGR starting from
\begin{equation}
    h_{A_1A_2A_3}=2\frac{\PPb_{12}^2\beta_1^2}{\beta_2^2\beta_3^2}\delta_{\lambda_1,-2}\delta_{\lambda_2,2}\delta_{\lambda_3,2}\delta\Big(\sum_i q_i\Big)\,,
\end{equation}
then the structure constant is
\begin{equation}
\fB_{A_1A_2A_3}=\frac{\PPb_{12}\beta_1}{\beta_2\beta_3}\big(\delta_{\lambda_1,-2}\delta_{\lambda_2,2}\delta_{\lambda_3,2}+\frac{\beta_2^4}{\beta_1^4}\delta_{\lambda_1,2}\delta_{\lambda_2,-2}\delta_{\lambda_3,2}+\frac{\beta_3^4}{\beta_1^4}\delta_{\lambda_1,2}\delta_{\lambda_2,2}\delta_{\lambda_3,-2}\big)\delta\Big(\sum_i q_i\Big)\,.
\end{equation}
The Jacobi identity is then verified thanks to the identity
\begin{equation}
    \PPb_{12}\PPb_{34}+\PPb_{23}\PPb_{14}+\PPb_{31}\PPb_{24}=A^2-B^2+C^2-A^2+B^2-C^2=0\,,
\end{equation}
that on-shell coincides with a special case of the Schouten identity. In \cite{Ponomarev:2016lrm,Serrani:2025owx} it was shown that a theory with the following couplings $C^{2,-2,2}=C^{\lambda,-\lambda,2}=\ell$ is consistent. Therefore, we could search for its gauge algebra:
\begin{align}
\begin{split}
\fB_{A_1A_2A_3}=&\,\ell\Big(\frac{\PPb_{12}\beta_1}{\beta_2\beta_3}(\delta_{\lambda_1,-2}\delta_{\lambda_2,2}\delta_{\lambda_3,2}+\left(\frac{\beta_2}{\beta_1}\right)^4\delta_{\lambda_1,2}\delta_{\lambda_2,-2}\delta_{\lambda_3,2}+\left(\frac{\beta_3}{\beta_1}\right)^4\delta_{\lambda_1,2}\delta_{\lambda_2,2}\delta_{\lambda_3,-2})\\
&+ \frac{\PPb_{12}\beta_1^{\lambda-1}}{\beta_2^{\lambda-1}\beta_3}(\delta_{\lambda_1,-\lambda}\delta_{\lambda_2,\lambda}\delta_{\lambda_3,2}+\left(\frac{\beta_2}{\beta_1}\right)^{2\lambda-2}\delta_{\lambda_1,\lambda}\delta_{\lambda_2,-\lambda}\delta_{\lambda_3,2})\\
&+\frac{\PPb_{12}\beta_1^{\lambda-1}}{\beta_3^{\lambda-1}\beta_2}(\delta_{\lambda_1,-\lambda}\delta_{\lambda_2,2}\delta_{\lambda_3,\lambda}+\left(\frac{\beta_3}{\beta_1}\right)^{2\lambda-2}\delta_{\lambda_1,\lambda}\delta_{\lambda_2,2}\delta_{\lambda_3,-\lambda})\\
&+\frac{\PPb_{12}\beta_2^{\lambda-1}}{\beta_3^{\lambda-1}\beta_1}(\delta_{\lambda_1,2}\delta_{\lambda_2,-\lambda}\delta_{\lambda_3,\lambda}+\left(\frac{\beta_3}{\beta_2}\right)^{2\lambda-2}\delta_{\lambda_1,2}\delta_{\lambda_2,\lambda}\delta_{\lambda_3,-\lambda})\Big)\delta\Big(\sum_i q_i\Big)\,,
\end{split}
\end{align}
and also here, the Jacobi identity is then verified thanks to the identity
\begin{equation}\label{JI_gauge_algebra}
    \PPb_{12}\PPb_{34}+\PPb_{23}\PPb_{14}+\PPb_{31}\PPb_{24}=A^2-B^2+C^2-A^2+B^2-C^2=0\,.
\end{equation}
Now that we understood the general idea, we can try to solve the Jacobi identity for a general gauge algebra.

\subsection{Jacobi identity and celestial OPE associativity}
The Jacobi identity for a generic gauge algebra is
\begin{equation}
    \fB_{A_1A_2B}\fB^B_{\;\;A_3A_4}+\fB_{A_2A_3B}\fB^B_{\;\;A_1A_4}+\fB_{A_3A_1B}\fB^B_{\;\;A_2A_4}=0\,.
\end{equation}
Using the definition \eqref{structure_const} we get
\begin{align}\label{JI}
\begin{split}
\sum_{\lambda_i}(-)^{\lambda_i}\Big(\mathcal{C}^{1234\lambda_i}\PPb_{12}^{\lambda_{12}+\lambda_i-1}&\PPb_{34}^{\lambda_{34}-\lambda_i-1}+\mathcal{C}^{2314\lambda_i}\PPb_{23}^{\lambda_{23}+\lambda_i-1}\PPb_{14}^{\lambda_{14}-\lambda_i-1}\\
&+\mathcal{C}^{3124\lambda_i}\PPb_{31}^{\lambda_{13}+\lambda_i-1}\PPb_{24}^{\lambda_{24}-\lambda_i-1}\Big)=0\,,
\end{split}
\end{align}
and it coincides with the OPE associativity constraint \eqref{OPE_Ass_LC}! Similar observations had already been made in \cite{Guevara:2022qnm,Guevara:2021abz}.

If we assume that the cubic vertices used to construct the gauge algebra belong to some representation of a gauge group $G$, and then, for example, we assume they are accompanied by the structure constants $f_{abc}$, we get 
\begin{align}\notag
\sum_{\lambda_i}(-)^{\lambda_i}\Big(\mathcal{C}^{1234\lambda_i}\PPb_{12}^{\lambda_{12}+\lambda_i-1}&\PPb_{34}^{\lambda_{34}-\lambda_i-1}f_{a_1a_2c}f^c_{\;\;a_3a_4}+\mathcal{C}^{2314\lambda_i}\PPb_{23}^{\lambda_{23}+\lambda_i-1}\PPb_{14}^{\lambda_{14}-\lambda_i-1}f_{a_2a_3c}f^c_{\;\;a_1a_4}+\\
&+\mathcal{C}^{3124\lambda_i}\PPb_{31}^{\lambda_{13}+\lambda_i-1}\PPb_{24}^{\lambda_{24}-\lambda_i-1}f_{a_3a_1c}f^c_{\;\;a_2a_4}\Big)=0\,.
\end{align}
Now using the Jacobi identity for $f_{abc}$, we obtain two independent constraints
\begin{align}\label{JI_colour}
&\sum_{\lambda_i}(-)^{\lambda_i}\Big(\mathcal{C}^{1234\lambda_i}\PPb_{12}^{\lambda_{12}+\lambda_i-1}\PPb_{34}^{\lambda_{34}-\lambda_i-1}+\mathcal{C}^{2314\lambda_i}\PPb_{23}^{\lambda_{23}+\lambda_i-1}\PPb_{14}^{\lambda_{14}-\lambda_i-1}\Big)=0\,,\\
&\sum_{\lambda_i}(-)^{\lambda_i}\Big(\mathcal{C}^{1234\lambda_i}\PPb_{12}^{\lambda_{12}+\lambda_i-1}\PPb_{34}^{\lambda_{34}-\lambda_i-1}+\mathcal{C}^{3124\lambda_i}\PPb_{31}^{\lambda_{13}+\lambda_i-1}\PPb_{24}^{\lambda_{24}-\lambda_i-1}\Big)=0\,,
\end{align}
which after moving the external helicity using the symmetries \eqref{coupling_sym}, the first corresponds to the colour ordering $[1234]$ and then to \eqref{OPE_Ass_color}, while the second to the colour ordering $[1243]$. Again, a term $\theta_{\lambda_i}=(-)^{\lambda_i}$ can be added to match notations with the OPE associativity case. Moreover, by assuming a more general gauge algebra including all the possible generic couplings \eqref{cubic_vertex_general} contributing, we get the same expression as in Appendix \ref{Paper2_AppendixD}.

\subsection{Jacobi identity and vanishing of the four-point amplitude}
Another interesting observation is the relation between the Jacobi identity and the vanishing of the four-point tree-level amplitude for a generic chiral theory. Using identities from Appendix \ref{Paper2_AppendixC} valid on-shell, the four-point amplitude takes the form
\begin{equation}
\begin{aligned}\label{4_pt_amplitude}
\mathcal{A}&=\mathcal{A}_s+\mathcal{A}_t+\mathcal{A}_u=
    \sum_{\lambda_i}(-)^{\lambda_i}\mathcal{C}^{1234\lambda_i}\frac{\PPb_{12}^{\lambda_{12}+\lambda_i}}{\beta_1^{\lambda_1}\beta_2^{\lambda_2}}\frac{1}{(q_1+q_2)^2}\frac{\PPb_{34}^{\lambda_{34}-\lambda_i}}{\beta_3^{\lambda_3}\beta_4^{\lambda_4}}+2\leftrightarrow 4+2\leftrightarrow 3\\
    &=\frac{\PPb_{12}\PPb_{34}}{(q_1+q_2)^2\prod_{i=1}^4\beta_i^{\lambda_i}}\sum_{\lambda_i}(-)^{\lambda_i}\Big(\mathcal{C}^{1234\lambda_i}\PPb_{12}^{\lambda_{12}+\lambda_i-1}\PPb_{34}^{\lambda_{34}-\lambda_i-1}+\mathcal{C}^{1423\lambda_i}\PPb_{23}^{\lambda_{23}+\lambda_i-1}\PPb_{14}^{\lambda_{14}-\lambda_i-1}\\
    &\qquad\qquad\qquad\qquad\qquad\qquad\qquad+\mathcal{C}^{3124\lambda_i}\PPb_{31}^{\lambda_{13}+\lambda_i-1}\PPb_{24}^{\lambda_{24}-\lambda_i-1}\Big)=0\,.
\end{aligned}
\end{equation}
Therefore, the vanishing of the four-point amplitude for a chiral theory gives the same expression as the Jacobi identity \eqref{JI} and the OPE associativity \eqref{OPE_Ass_LC}.

In the presence of colours, the four-point tree-level amplitude  can be written in terms of the colour-ordered ones \eqref{colour_ordered_ampl} and by considering a single colour-ordered amplitude, here $[1234]$, since the same is valid for the others, we obtain
\begin{align}
    \begin{split}
    \tilde{\mathcal{A}}(1234)=\frac{\PPb_{12}\PPb_{34}}{(q_1+q_2)^2\prod_{i=1}^4\beta_i^{\lambda_i}}&\sum_{\lambda_i}(-)^{\lambda_i}\Big(\mathcal{C}^{1234\lambda_i}\PPb_{12}^{\lambda_{12}+\lambda_i-1}\PPb_{34}^{\lambda_{34}-\lambda_i-1}\\
    &-\,\mathcal{C}^{2341\lambda_i}\PPb_{23}^{\lambda_{23}+\lambda_i-1}\PPb_{41}^{\lambda_{41}-\lambda_i-1}\Big)=0\,,
    \end{split}
\end{align}
where we can always introduce the $\theta_{\lambda_i}$ factors and correspond to the Jacobi identity for the gauge algebra in \eqref{JI_colour}, and to the OPE associativity in \eqref{OPE_Ass_color}. Moreover, by considering even more generic couplings \eqref{cubic_vertex_general} contributing to the amplitude, we get the same expression as in Appendix \ref{Paper2_AppendixD} with the same prefactor as in the expressions above.

\subsection{Relation to the light-cone quartic holomorphic constraint}
We revisit the statement made in \cite{Ponomarev:2017nrr} about the relation between the light-cone consistency condition and the Jacobi identity for the gauge algebra. We point out that, on the energy-shell,\footnote{That is, we impose the energy conservation (i.e. with $\mathcal{H}=0$) for the external fields.} the two conditions are the same up to a specific product of couplings, and we elaborate on the implications of such a difference.

Let us recall the form of the total free Hamiltonian $\mathcal{H}$ and the boost operator $\mathcal{J}$:
\begin{align}
    &\mathcal{H}=\sum_{i=0}^nh_2^{\lambda_i}(q_i)=\sum_{i=0}^n-\frac{q_i\bar{q}_i}{\beta_i}\,,&
    &\mathcal{J}=\sum_{i=0}^nj_2^{\lambda_i}(q_i)=\sum_{i=0}^n\left(-\frac{q_i\bar{q}_i}{\beta_i}\frac{\partial}{\partial\bar{q}_i}-q_i\frac{\partial}{\partial\beta_i}+\lambda_i\frac{q_i}{\beta_i}\right)\,.
\end{align}
The light-cone (holomorphic) constraint is equivalent to the Lorentz invariance of the S-matrix. This can be rewritten as a set of Ward identities for all $n$-point off-shell amplitudes \cite{Ponomarev:2016cwi}.\footnote{For a definition of the off-shell amplitudes, we refer to \cite{Ponomarev:2016cwi}.} The statement is the following:
\begin{equation}
    [H,J]=0\qquad
    \Longleftrightarrow\qquad
    [\mathcal{A},J_2]=0\,,
\end{equation}
where $\mathcal{A}$ is the off-shell amplitude.
In particular, for chiral theories, where only cubic couplings are present, we can focus on the four-point amplitude\footnote{Indeed, the quartic holomorphic constraint is the only constraint for a cubic theory. It is equivalent to the Poincaré invariance of the four-point amplitude. Iteratively, one can show that higher-point amplitudes have vanishing cuts (e.g. a cut of a five-point amplitude is the product of the (vanishing) four-point and a three-point), i.e. they should vanish as well.}
\begin{equation}\label{H3J3=A4J2}
    [H_3,J_3]=0\qquad
    \Longleftrightarrow\qquad
    [\mathcal{A}_4,J_2]=0\,.
\end{equation}
We can rewrite the condition above as follows
\begin{equation}\label{Amplitude4}
\delta\left(\sum_{i=1}^4q_i\right)\mathcal{J}\left(\sum_{\lambda_i}h_3^{\lambda_1,\lambda_2,\lambda_i}(q_1,q_2,q_{\lambda_i})\frac{1}{s}h_3^{\lambda_3,\lambda_4,-\lambda_i}(q_3,q_4,-q_{\lambda_i})+ (t,u)\text{-channels}\right)=0\,,
\end{equation}
where $\lambda_i$ represents the helicities of the particles exchanged in the quartic tree-level diagram. The equivalence \eqref{H3J3=A4J2} may be established by a direct computation and arises as a consequence of the following identity:
\begin{align}
\begin{split}
&\mathcal{J}\left(h_3^{\lambda_1,\lambda_2,\lambda_i}(q_1,q_2,q_{\lambda_i})\frac{1}{s}h_3^{\lambda_3,\lambda_4,-\lambda_i}(q_3,q_4,-q_{\lambda_i})\right)=\\
&\left((-)^{\lambda_i}\frac{(\lambda_1+\lambda_i-\lambda_2)\beta_1-(\lambda_2+\lambda_i-\lambda_1)\beta_2}{\beta_1+\beta_2}\PPb_{12}^{\lambda_{12}+\lambda_i-1}\PPb_{34}^{\lambda_{34}-\lambda_i}+(1,2)\leftrightarrow (3,4)\right)\,,
\end{split}
\end{align}
where $s=\tfrac{1}{2}([12]\langle 12\rangle+[34]\langle 34\rangle)$, and the use of momentum conservation and energy conservation (i.e. with $\mathcal{H}=0$) for the external fields is required. As we have reviewed above, and following \cite{Ponomarev:2016cwi}, to isolate the gauge algebra structure, we can factor out the SDYM kinematical part of the vertices as
\begin{align}
    &h_3^{\lambda_1,\lambda_2,\lambda_i}=\fB_{A_1A_2B}h^{1,1,-1}_\text{YM}\,,&
    &h_3^{\lambda_3,\lambda_4,-\lambda_i}=\fB^B_{\phantom{B}A_3A_4}h^{1,-1,1}_\text{YM}\,.
\end{align}
Then, we rewrite \eqref{Amplitude4} in the following form:
\begin{align}
    \begin{split}
    \delta\left(\sum_{i=1}^4q_i\right)\Bigg(&\Big(h^{1,1,-1}_\text{YM}\frac{1}{s}h^{1,-1,1}_\text{YM}\Big)\mathcal{J}^{\lambda_1-1,\lambda_2-1,\lambda_3-1,\lambda_4+1}[\fB_{A_1A_2B}\fB^B_{\phantom{B}A_3A_4}]\\
    &+\mathcal{J}^{1,1,1,-1}[h^{1,1,-1}_\text{YM}\frac{1}{s}h^{1,-1,1}_\text{YM}](\fB_{A_1A_2B}\fB^B_{\;\;A_3A_4})+(t,u)\text{-channels}\Bigg)\,.
    \end{split}
\end{align}
By introducing the following operator
\begin{align}
    \mathcal{O}_s=\Big(h^{1,1,-1}_\text{YM}\frac{1}{s}h^{1,-1,1}_\text{YM}\Big)\mathcal{J}^{\lambda_1-1,\lambda_2-1,\lambda_3-1,\lambda_4+1}+\mathcal{J}^{1,1,1,-1}[h^{1,1,-1}_\text{YM}\frac{1}{s}h^{1,-1,1}_\text{YM}]\,,
\end{align}
and the corresponding one for the $t$ and $u$ channels. The equation above has the following schematic form:
\begin{equation}\label{Jacobi_O}
    \mathcal{O}_s[\fB_{A_1A_2B}\fB^B_{\phantom{B}A_3A_4}]+\mathcal{O}_t[\fB_{A_2A_3B}\fB^B_{\;\;A_1A_4}]+\mathcal{O}_u[\fB_{A_3A_1B}\fB^B_{\phantom{B}A_2A_4}]=0\,.
\end{equation}
The idea of \cite{Ponomarev:2016cwi} was to show that
\begin{equation}
    \mathcal{O}=\mathcal{O}_s=\mathcal{O}_t=\mathcal{O}_u\,.
\end{equation}
By employing momentum conservation inside the Jacobi identity, this can be proven for $\mathcal{H}=0$. Indeed, the three operators will differ only by terms proportional to the total free Hamiltonian $\mathcal{H}$.

We have thus established that any solution of the Jacobi identity must also satisfy the light-cone holomorphic constraint. However, proving the converse requires first determining the kernel of the operator $\mathcal{J}$. To this end, we apply $\mathcal{J}$ to the most general chiral polynomial function $K(\PPb_{12},\PPb_{34},\beta_1,\beta_2,\beta_3,\beta_4)$ and we obtain
\begin{align}
    &\mathcal{J}^{\lambda_1,\lambda_2,\lambda_3,\lambda_4}\cdot K=0\,,&
    &K=\beta_1^{\lambda_1-n}\beta_2^{\lambda_2-n}\beta_3^{\lambda_3-m}\beta_4^{\lambda_4-m}\PPb_{12}^n\PPb_{34}^m\,.
\end{align}
This can be translated to the kernel of $\mathcal{O}_s$, $\mathcal{O}_t$, and $\mathcal{O}_u$ and gives
\begin{subequations}
\begin{align}
    &\mathcal{O}_s\cdot K_s=0\,,&
    &K_s=\beta_1^{\lambda_1-n}\beta_2^{\lambda_2-n}\beta_3^{\lambda_3-m}\beta_4^{\lambda_4-m-2}\PPb_{12}^n\PPb_{34}^m\,,\\
    &\mathcal{O}_t\cdot K_t=0\,,&
    &K_t=\beta_2^{\lambda_2-n}\beta_3^{\lambda_3-n}\beta_1^{\lambda_1-m}\beta_4^{\lambda_4-m-2}\PPb_{23}^n\PPb_{14}^m\,,\\
    &\mathcal{O}_u\cdot K_u=0\,,&
    &K_u=\beta_3^{\lambda_3-n}\beta_1^{\lambda_1-n}\beta_2^{\lambda_2-m}\beta_4^{\lambda_4-m-2}\PPb_{31}^n\PPb_{24}^m\,.
\end{align}
\end{subequations}
At this point, we can look if the operators in \eqref{Jacobi_O} contain terms in the kernel of $\mathcal{O}_s$, $\mathcal{O}_t$, and $\mathcal{O}_u$. The explicit form of the product of two structure constants is
\begin{equation}
    \fB_{A_1A_2B}\fB^B_{\phantom{B}A_3A_4}\sim C^{\lambda_1,\lambda_2,\lambda_i}C^{-\lambda_i,\lambda_3,\lambda_4}\frac{\PPb_{12}^{\lambda_{12}+\lambda_i-1}\PPb_{34}^{\lambda_{34}-\lambda_i-1}}{\beta_1^{\lambda_1-1}\beta_2^{\lambda_2-1}\beta_3^{\lambda_3-1}\beta_4^{\lambda_4+1}}\,.
\end{equation}
This term is in the kernel of the operator $\mathcal{O}_s$ only for the specific case of $C^{\lambda_1,\lambda_1,0}C^{0,\lambda_2,\lambda_2}$, where
\begin{equation}
    \mathcal{O}_s\cdot \frac{\PPb_{12}^{2\lambda_1-1}\PPb_{34}^{2\lambda_2-1}}{\beta_1^{\lambda_1-1}\beta_2^{\lambda_1-1}\beta_3^{\lambda_2-1}\beta_4^{\lambda_2+1}}=0\,,
\end{equation}
and the same, with obvious modifications, holds for $\mathcal{O}_t$ and $\mathcal{O}_u$.
These products of couplings coincide with those left unconstrained by the quartic holomorphic constraint, as shown in \cite{Serrani:2025owx}. We then identified the transformation relating the two constraints as that induced by the operators $\mathcal{O}_s$, $\mathcal{O}_t$, and $\mathcal{O}_u$. 

To conclude, we showed that the solutions to the quartic holomorphic constraint and to the OPE associativity constraint coincide, up to the products of couplings $C^{\lambda_1,\lambda_1,0}C^{0,\lambda_2,\lambda_2}$ which remain unconstrained in the holomorphic constraint.\footnote{Notice that this can also be checked directly by solving both constraints. The solution to the OPE associativity found here, and the solution to the holomorphic constraint in \cite{Serrani:2025owx}, indeed coincide, up to the special product $C^{\lambda_1,\lambda_1,0}C^{0,\lambda_2,\lambda_2}$.}

This distinction is significant, as already emphasised in \cite{Serrani:2025owx}. In particular, it not only allows for terms such as $\sim F^3$ and $\sim R^3$ to appear with free coupling constants --- consistent with the covariant formulation of lower-spin theories --- but also permits non-vanishing amplitudes. Indeed, even in chiral theories and with higher-spin interactions, where the Weinberg low-energy theorem would suggest a vanishing amplitude, this argument does not apply to abelian terms (also referred to as Born–Infeld type interactions), which can give rise to non-vanishing contributions.

\section{Conclusions}\label{Paper2_section7}

After a review of both the light-cone quartic holomorphic constraint in $4d$ higher-spin theories \cite{Ponomarev:2016lrm,Metsaev:1991mt,Metsaev:1991nb}, and the OPE associativity in $2d$ Celestial CFT \cite{Ren:2022sws,Mago:2021wje}, we have solved the latter using the same ideas employed to solve the former \cite{Serrani:2025owx}. We have also clarified their relation.

This is inspired by \cite{Ren:2022sws,Monteiro:2022lwm}, where it was argued that the Metsaev couplings of the full chiral higher-spin theory are solutions to the OPE associativity constraint. In this work, we established an explicit relation between the two constraints. We identify the operator that relates the two and provide a method to classify all solutions, including those involving higher-spin fields and higher-derivative vertices. Moreover, we also uncover an interesting relation with the vanishing of the scattering amplitude and the Jacobi identity of a gauge algebra  constructed from the cubic vertices, as discussed in \cite{Ponomarev:2017nrr}. 

In \cite{Serrani:2025owx}, we initiated the programme of classifying all possible chiral higher-spin theories, which, for lower-derivative interactions, also leads to an OPE associative Celestial CFTs. Differences can arise for higher-derivative theories due to the unconstrained product $C^{\lambda_1,\lambda_1,0}C^{0,\lambda_2,\lambda_2}$, as we showed for certain low-spin examples.

It would be interesting to provide a geometrical interpretation of the gauge algebras constructed from the holomorphic vertices \eqref{structure_const}, which give rise to OPE-associative celestial CFTs. A partial step was taken in \cite{Ponomarev:2017nrr}, where the gauge algebra of the full chiral higher-spin theory and its truncations to HS-SDYM and HS-SDGR were described. In particular, dual celestial chiral algebras exist for SDYM and SDGR \cite{Monteiro:2022lwm}. Extending this construction to all OPE-associative celestial CFTs would be natural, since we have shown that each bulk gauge algebra gives rise to an associative OPE on the boundary. Consequently, OPE associativity guarantees the existence of an underlying celestial chiral algebra, which could also admit a twistor interpretation, as in SDGR \cite{Adamo:2021lrv}.

As shown in \cite{Ponomarev:2017nrr}, with the help of the gauge algebra, one can reformulate the equations of motion of the Chiral higher-spin gravities as the self-duality constraint. Therefore, one can also refer to Chiral higher-spin gravities as self-dual ones. It is then tempting to place self-duality at the centre of the picture and interpret the OPE associativity, the vanishing of tree-level amplitudes, and the holomorphic light-cone constraint as consequences of self-duality. 

A natural question that arises is whether some of the results of the paper can be extended to the full quartic level by identifying the complete quartic constraint with the OPE associativity constraint of the “all-order” celestial OPE \cite{Adamo:2022wjo,Ren:2023trv}. More broadly, it would be worthwhile to extend this correspondence in two directions: first, by incorporating the classification of massive cubic vertices \cite{Metsaev:2005ar}, and second, by investigating how the relation might persist --- or be modified --- at the loop level. Loop corrections for  Chiral higher-spin gravity were studied in \cite{Skvortsov:2018jea,Skvortsov:2020wtf,Skvortsov:2020gpn,Tsulaia:2022csz}. It would be interesting to see what happens to the theories \cite{Serrani:2025owx} with finitely many higher-spin fields, where some of the results of \cite{Skvortsov:2018jea,Skvortsov:2020wtf,Skvortsov:2020gpn,Tsulaia:2022csz} do not apply directly.

Another important direction for future work is the classification of higher-derivative cases, both for the OPE associativity and for the light-cone holomorphic constraint. Here, the relation to twistor methods should be fruitful \cite{Tran:2021ukl,Herfray:2022prf,Tran:2022tft,Adamo:2022lah,Mason:2025pbz,Tran:2025xbt,Tran:2025uad}. It would also be important to clarify the relation to the results of \cite{Ponomarev:2022atv,Ponomarev:2022ryp,Ponomarev:2022qkx} where a ``flat space'' analogue of the singleton representation was proposed and amplitudes of Chiral higher-spin gravity were computed.

At least the theories of HS-SDYM type \cite{Ponomarev:2016lrm,Ponomarev:2017nrr,Krasnov:2021nsq,Monteiro:2022xwq,Serrani:2025owx}, fully classified in \cite{Serrani:2025owx}, are conformally invariant. Therefore, none of the main conclusions of the paper should change once a conformal transformation is applied to map them to anti-de Sitter space. It remains to be seen what role the celestial OPE would play in AdS/CFT correspondence (see \cite{Skvortsov:2018uru,Sharapov:2022awp,Jain:2024bza,Aharony:2024nqs} for the discussion of Chiral higher-spin gravity in the AdS/CFT context) and whether it can be extended beyond the theories of (HS)-SDYM type that are conformal, where covariant formulations \cite{Sharapov:2022faa,Sharapov:2022wpz,Sharapov:2022awp,Sharapov:2022nps,Sharapov:2023erv,Skvortsov:2024rng,Tran:2025yzd} of higher-spin theories should play an important role. 

We conclude by noting that the holomorphic OPE in CCFT can be determined using Poincaré invariance \cite{Himwich:2021dau}, which also allows fixing the cubic vertices in the bulk, interpreted as OPE coefficients in the CCFT. This parallels the role of the cubic light-cone constraint. In contrast, OPE associativity appears unrelated to Poincaré invariance. However, its close connection to the quartic holomorphic light-cone constraint in the bulk suggests otherwise. The slight discrepancy --- namely, the fact that the product $C^{\lambda_1,\lambda_1,0}C^{0,\lambda_2,\lambda_2}$ remains unconstrained by the light-cone holomorphic constraint --- deserves further investigation and may have a deeper explanation.
We hope to clarify these aspects in future work by trying to extend this correspondence to the full quartic level.

\section*{Acknowledgments}
\label{sec:Aknowledgements1}
This project has received funding from the European Research Council (ERC) under the European Union’s Horizon 2020 research and innovation programme (grant agreement No 101002551). I am grateful to Evgeny Skvortsov for suggesting to elaborate on \cite{Ren:2022sws} and for many useful discussions. I am grateful to Akshay Yelleshpur Srikant and Dmitry Ponomarev for useful comments on the draft. I thank the anonymous Referee for the valuable suggestions and insightful questions.

\appendix

\section{Light-cone notations}\label{Paper2_AppendixA}

We use both light-cone coordinates and the light-cone gauge (in fact, double-null).
In flat spacetime, we adopt the $4d$ Minkowski metric with ``mostly plus'' signature
\begin{align}
    &x^{\mu}=(x^0,x^1,x^2,x^3)\,,&
    &\eta^{\mu\nu}=\text{diag}(-,+,+,+)\,,\\
    &ds^2=-(dx^0)^2+(dx^1)^2+(dx^2)^2+(dx^3)^2\,.
\end{align}
We define the light-cone coordinates as
\begin{align}
    x^+&=\frac{x^3+x^0}{\sqrt{2}}\,,&
    x^-&=\frac{x^3-x^0}{\sqrt{2}}\,,&
    x&=\frac{x^1-ix^2}{\sqrt{2}}\,,&
    \bar{x}&=\frac{x^1+ix^2}{\sqrt{2}}\,,\\
    \partial_-=\partial^+ &= \frac{\partial^3+\partial^0}{\sqrt{2}}\,, &
    \partial_+=\partial^- &= \frac{\partial^3-\partial^0}{\sqrt{2}}\,, &
    \partial &= \frac{\partial^1+i\partial^2}{\sqrt{2}}\,, &
    \bar{\partial} &= \frac{\partial^1-i\partial^2}{\sqrt{2}}\,,
\end{align}
where the metric and the wave-operator become
\begin{align}
    &x^{\mu}=(x^+,x^-,x,\bar{x})\,,&
    &ds^2=2\,dx^+ dx^- + 2\,dx d\bar{x}\,,\\
    &\Box=\partial_{\mu}\partial^{\mu}=2(\partial^+\partial^-+\partial\bar{\partial})\,,
\end{align}
the light-front scalar product becomes
\begin{align}
    A_{\mu}B^{\mu}=A_+B^++A_-B^-+A\bar{B}+\bar{A}B=A^-B^++A^+B^-+A\bar{B}+\bar{A}B\,,
\end{align}
and our definitions for the derivatives imply
\begin{align}
    \partial^+ x^-=\partial^-x^+=\partial x=\bar{\partial}\bar{x}=1\,.
\end{align}
In the light-front, $x^+$ is taken to be the time variable, and $\partial^-$ is the time derivative. Moreover, one assumes that $\partial^+$ is always non-zero and therefore can be inverted as an operator.

\section{Celestial CFT notations for massless fields}\label{Paper2_AppendixB}
In CCFT, to capture outgoing radiative data near future null infinity $\mathscr{I}^+$, we parametrise massless momenta (i.e. $q^2=0$) as 
\begin{align}
    \begin{split}
     q^{\mu}(\omega,z,\bar{z},\epsilon)&=\epsilon\,\omega(1+z\bar{z},z+\bar{z},i(\bar{z}-z),1-z\bar{z})\\
     &=\epsilon\,\omega(1+|z|^2,2\,\text{Re}(z),2\,\text{Im}(z),1-|z|^2)\,,
     \end{split}
\end{align}
where $\epsilon=\pm$ distinguishes outgoing ($+$) and incoming\footnote{We use the all-outgoing convention for amplitudes; then all particles are considered to be outgoing, and a minus sign is introduced for incoming particles.} ($-$) particles. The energy of a massless particle is given by $q^0=\omega(1+z\bar{z})$, with $\omega>0$. The complex variables $z$ and $\bar{z}$ serve as coordinates on the celestial sphere. While they are independent in complexified Minkowski space, under the Lorentzian signature with the ``mostly plus'' convention $\eta_{\mu\nu}=\text{diag}(-,+,+,+)$, they are taken to be complex conjugates of each other.

A scattering process involving $n$ massless fields with momenta $q^{\mu}_i$ and spins $s_i$ can, once expressed in a basis of boost eigenstates, be reinterpreted as a correlation function of $n$ conformal primary operators in the dual $2d$ CCFT. Each operator is characterised by conformal weights $(h_i,\bar{h}_i)$, from which one can obtain the spin $s_i=h_i-\bar{h}_i$ and the conformal dimension $\Delta_i=h_i+\bar{h}_i$.

\section{Spinor-helicity in both formalism}\label{Paper2_AppendixC}
For our spinor-helicity conventions for massless particles, we follow \cite{Elvang:2013cua}:
\begin{align}
    &q_{a\dot{b}}= q_{\mu}(\sigma^{\mu})_{a\dot{b}}\,,&
    &\text{det}(q_{a\dot{b}})=-q^{\mu}q_{\mu}=m^2\,,&
    &q^2=0\;\; \Rightarrow\;\; q_{a\dot{b}}=-|q]_a\langle q|_{\dot{b}}\equiv -\lambda_a\tilde{\lambda}_{\dot{b}}\,,
\end{align}
\begin{align}
        &\langle ij\rangle\definition \langle q_i|_{\dot{a}}|q_j\rangle^{\dot{a}}\equiv\tilde{\lambda}_{i\dot{a}}\tilde{\lambda}_j^{\dot{a}}\,,&
    &[ij]\definition [q_i|^a|q_j]_a\equiv \lambda_i^{a}\lambda_{ja}\,\,,&
    &\lambda^{a}=\epsilon^{ab}\lambda_{b}\,,&
&\tilde{\lambda}^{\dot{a}}=\epsilon^{\dot{a}\dot{b}}\tilde{\lambda}_{\dot{b}}\,,
\end{align}
\begin{align}
&\sigma^0=
    \begin{pmatrix}
        1 & 0\\
        0 & 1
    \end{pmatrix}\,,&
    &\sigma^1=
    \begin{pmatrix}
        0 & 1\\
        1 & 0
    \end{pmatrix}\,,&
    &\sigma^2=
    \begin{pmatrix}
        0 & -i\\
        i & 0
    \end{pmatrix}\,,&
    &\sigma^3=
    \begin{pmatrix}
        1 & 0\\
        0 & -1
    \end{pmatrix}\,,
\end{align}
\begin{align}
    &\epsilon^{ab}=-\epsilon_{ab}=
    \begin{pmatrix}
        0 & 1\\
        -1 & 0
    \end{pmatrix}\,,&
    &\epsilon^{\dot{a}\dot{b}}=-\epsilon_{\dot{a}\dot{b}}=
    \begin{pmatrix}
        0 & 1\\
        -1 & 0
    \end{pmatrix}\,.
\end{align}
Notice that for complex momenta $q^{\mu}$ the two spinors $(\lambda_a,\tilde{\lambda}_{\dot{b}})$ are independent two-dimensional complex vectors. In Minkowski space and for real momenta $q_{a\dot{b}}$ is hermitian, and the two spinors become complex conjugate $\tilde{\lambda}_{\dot{a}}=\pm(\lambda_a)^*$ (where the sign depends on whether the energy is taken to be positive or negative, then on the convention we use on the background flat metric).

Spinor-helicity variables $(\lambda_i,\tilde{\lambda}_i)$ are defined up to little group scaling $(\lambda_i,\tilde{\lambda}_i)\sim (t_i\lambda_i,t_i^{-1}\tilde{\lambda}_i)$ for $t_i\in\mathbb{C}^*$. In CCFT, we can choose the following adapted coordinates for massless particles, following \cite{Himwich:2021dau}:
\begin{align}
    &|q_i]_a\equiv\lambda_{ai}=\sqrt{\omega_i}\begin{pmatrix}
        -\bar{z}_i\\
        1
    \end{pmatrix}\,,&
    &\langle q_i|_{\dot{a}}\equiv\tilde{\lambda}_{\dot{a}i}=\epsilon_i\sqrt{\omega_i}\begin{pmatrix}
         -z_i & 1
    \end{pmatrix}\,,&
    &z_i,\bar{z}_i\in\mathbb{C}\,,&
    &\omega_i\in\mathbb{C}^*\,.
    \end{align}
Here, $(z_i,\bar{z}_i)$ are generally taken to be complex and independent of each other,\footnote{They become complex conjugates in Lorentzian signature, where they act as coordinates in the celestial sphere $\mathbb{C}\PP^1$.} $\omega_i$ specifies the energy of the particle and $\epsilon_i=\pm 1$ stands for outgoing $(+)$ and incoming $(-)$ particles. In the following, we consider all-outgoing particles, so we fix $\epsilon_i\cdot\epsilon_j=1$. Therefore, we get the following:
\begin{align}
[ij]=
    \sqrt{\omega_i\omega_j}\bar{z}_{ij},\quad
    \langle ij \rangle=-\sqrt{\omega_i\omega_j}z_{ij},\quad
z_{ij}= z_i-z_j,\quad 
\bar{z}_{ij}=\bar{z}_i-\bar{z}_j\,.
\end{align}
In the context of the light-cone Hamiltonian approach for massless higher-spin fields, we adopt the following notation \cite{Ponomarev:2016cwi}:
\begin{equation}
    q_{a\dot{b}}=q_{\mu}(\sigma^{\mu})_{a\dot{b}}=\sqrt{2}
    \begin{pmatrix}
        q^- & \bar{q}\\
        q & - \beta
    \end{pmatrix}\approx\sqrt{2}
    \begin{pmatrix}
        -\frac{q\bar{q}}{\beta} & \bar{q}\\
        q & - \beta
    \end{pmatrix}= -|q]_a\langle q|_{\dot{b}}=-\lambda_a\tilde{\lambda}_{\dot{b}}\,,
\end{equation}
\begin{align}
&\tilde{\lambda}_{\dot{a}i}=\frac{2^{\frac{1}{4}}}{\sqrt{\beta_i}}\begin{pmatrix}
    q_i & -\beta_i
    \end{pmatrix}\,,&
    &\langle ij\rangle=-\sqrt{\frac{2}{\beta_i\beta_j}}\PP_{ij}\,,&
    &\lambda_{ai}=\frac{2^{\frac{1}{4}}}{\sqrt{\beta_i}}
    \begin{pmatrix}
       \bar{q}_i\\
        -\beta_i
    \end{pmatrix}\,,&
    &[ij]=\sqrt{\frac{2}{\beta_i\beta_j}}\PPb_{ij}\,.
\end{align}
The two formalisms are related by
\begin{equation}
    z_{ij}=\frac{\PP_{ij}}{\beta_i\beta_j},\quad
    \bar{z}_{ij}=\frac{\PPb_{ij}}{\beta_i\beta_j},\quad
    z_i=\frac{q_i}{\beta_i},\quad
    \bar{z}_i=\frac{\bar{q}_i}{\beta_i},\quad
    \sqrt{\omega_i}=-2^{\frac{1}{4}}\sqrt{\beta_i}\,.
\end{equation}
The relation between the cubic amplitude and the cubic Hamiltonian density follows:
\begin{align}\label{AmplitudeandHamiltonian_holo}
    \mathcal{A}_3&=C^{\lambda_1,\lambda_2,\lambda_3}[12]^{d_{12}} [23]^{d_{23}} [31]^{d_{31}}=C^{\lambda_1,\lambda_2,\lambda_3}\frac{\sqrt{2}^{\lambda_{123}}\PPb^{\lambda_{123}}}{\beta_1^{\lambda_1}\beta_2^{\lambda_2}\beta_3^{\lambda_3}}=\sqrt{2}^{\lambda_{123}}h_3\,,\\\label{AmplitudeandHamiltonian_antiholo}
    \bar{\mathcal{A}}_3&=\bar{C}^{-\lambda_1,-\lambda_2,-\lambda_3}\langle 12\rangle^{-d_{12}} \langle 23\rangle^{-d_{23}} \langle 31\rangle^{-d_{31}}=\bar{C}^{-\lambda_1,-\lambda_2,-\lambda_3}\frac{\sqrt{2}^{\lambda_{123}}\PP^{-\lambda_{123}}}{\beta_1^{-\lambda_1}\beta_2^{-\lambda_2}\beta_3^{-\lambda_3}}=\sqrt{2}^{\lambda_{123}}\bar{h}_3\,,
\end{align}
where $\mathcal{A}_3$ is valid for $\lambda_{123}>0$ and $\bar{\mathcal{A}}_3$ for $\lambda_{123}<0$, and we have defined
\begin{align}
    &d_{12}=\lambda_{12}-\lambda_3\,,&
    &d_{23}=\lambda_{23}-\lambda_1\,,&
    &d_{31}=\lambda_{31}-\lambda_2\,.
\end{align}
We also have some standard relations for the $n$-point scattering: 
\begin{align}
   &\langle ij\rangle [ij]=-\frac{2}{\beta_i\beta_j}\PP_{ij}\PPb_{ij}=2\,q_i \cdot q_j=(q_i+q_j)^2\,,\\
   &\sum^{n}_{j=1}q^{\mu}_j=0\quad\Rightarrow\quad\sum^{n}_{j=1}\langle ij\rangle[jk]=\sum^{n}_{j=1}\frac{\PP_{ij}\PPb_{jk}}{\beta_j}=0\,.
\end{align}
In particular, using the relation above for the four-point scattering, we find
\begin{align}
    &\frac{\PPb_{12}\PPb_{34}}{(q_1+q_2)^2}=\frac{\PPb_{31}\PPb_{24}}{(q_1+q_3)^2}=\frac{\PPb_{14}\PPb_{23}}{(q_1+q_4)^2}\,,&
    &s_{ij}=-(q_i+q_j)^2\,,
\end{align}
where $s=s_{12}$, $t=s_{14}$, $u=s_{13}$ are the standard Mandelstam variables for massless four-point scattering. 

The relations above, in the light-cone approach, hold only when the external particles are on-shell.  Indeed, one of the main differences between the cubic Hamiltonian density and the amplitudes in \eqref{AmplitudeandHamiltonian_holo}-\eqref{AmplitudeandHamiltonian_antiholo} is that the latter are inherently on-shell objects, while the former contains off-shell information.

\section{OPE associativity: most general solution}\label{Paper2_AppendixD}

It is straightforward to generalise the solution of the OPE associativity constraint to the most general cubic vertices \eqref{cubic_vertex_general}. We make use of the following definitions:
\begin{align}\label{definitions_2}
    \begin{split}
    &\mathcal{F}^{1234}\definition \fA\fdu{a_1a_2}{c}\fA_{c\,a_3a_4}=\delta^d_c\,\fA\fdu{a_1a_2}{c}\fA_{d\,a_3a_4}\,,\\
    &k_{+}^{1234}\definition(-)^{\lambda_{12}}\sum_{\lambda_i}(-)^{\lambda_i}\mathcal{C}^{1234\lambda_i}\,,\qquad
    k_{-}^{1234}\definition \sum_{\lambda_i}\mathcal{C}^{1234\lambda_i}\,,\\
    &f_-^{1234}(A,B)\definition \sum_{\lambda_i}\mathcal{C}^{1234\lambda_i}(A-B)^{\lambda_{12}+\lambda_i-1}(A+B)^{\lambda_{34}-\lambda_i-1}\,,
    \end{split}
\end{align}
where the upper index $1234$ denotes the order of the external helicities. Moreover, $\fA\fdu{a_1a_2}{c}\fA_{c\,a_3a_4}$ indicates the natural pairing between the positive and negative helicity fields $(\phi^{+\lambda})^a$ and $(\phi^{-\lambda})_b$, where the negative helicity fields take values in the dual vector space.\footnote{For example, if $f^a_{\phantom{a}bc}$ are the structure constants of some Lie algebra $\mathfrak{g}$, and $(\phi^{+\lambda})^a$ transforms in the adjoint representation, then $(\phi^{-1})_b$ transforms in the canonical dual to the adjoint, i.e. in the coadjoint one.} The contraction is performed with the standard Poisson bracket for fields decorated with vector space indices and reads
\begin{equation}
    [(\phi^{\lambda}_q)^a,(\phi^{s}_p)_b ]= \delta^{\lambda,-s}\delta^a_b\frac{\delta^3(q+p) }{2q^+}\,.
\end{equation}  
For general vertices, the holomorphic OPE associativity constraint takes the form
\begin{align}\label{OPE_Ass_general}
\nonumber
\sum_{\lambda_i}&\Big(\mathcal{C}^{1234\lambda_i}(\mathcal{F}^{1234}+\mathcal{F}^{2134}+\mathcal{F}^{1243}+\mathcal{F}^{2143})\PPb_{12}^{\lambda_{12}+\lambda_i-1}\PPb_{34}^{\lambda_{34}-\lambda_i-1}\\
&+\mathcal{C}^{3124\lambda_i}(\mathcal{F}^{3124}+\mathcal{F}^{1324}+\mathcal{F}^{3142}+\mathcal{F}^{1342})\PPb_{31}^{\lambda_{13}+\lambda_i-1}\PPb_{24}^{\lambda_{24}-\lambda_i-1}\\
\nonumber
&+\mathcal{C}^{1423\lambda_i}(\mathcal{F}^{1423}+\mathcal{F}^{4123}+\mathcal{F}^{1432}+\mathcal{F}^{4132})\PPb_{14}^{\lambda_{14}+\lambda_i-1}\PPb_{23}^{\lambda_{23}-\lambda_i-1}\Big)=0\,,
\end{align}
where we used the symmetry property \eqref{coupling_sym}, that of $\PPb$ and an additional minus sign due to the antisymmetry of $\langle 12\rangle$ and $\langle i3\rangle$ appearing in the denominators when computing the residue in \eqref{computing_residues}; each exchange of spinor labels introduces an extra minus sign.
Writing \eqref{OPE_Ass_general} in terms of the independent variables $A,B,C$, we find
\begin{align}
\begin{split}
\sum_{\lambda_i}&\Big(\mathcal{C}^{1234\lambda_i}(\mathcal{F}^{1234}+\mathcal{F}^{2134}+\mathcal{F}^{1243}+\mathcal{F}^{2143})(A-B)^{\lambda_{12}+\lambda_i-1}(A+B)^{\lambda_{34}-\lambda_i-1}\\
&+\mathcal{C}^{1342\lambda_i}(\mathcal{F}^{1342}+\mathcal{F}^{3142}+\mathcal{F}^{1324}+\mathcal{F}^{3124})(B-C)^{\lambda_{13}+\lambda_i-1}(B+C)^{\lambda_{24}-\lambda_i-1}\\
&+\mathcal{C}^{1423\lambda_i}(\mathcal{F}^{1423}+\mathcal{F}^{4123}+\mathcal{F}^{1432}+\mathcal{F}^{4132})(C-A)^{\lambda_{14}+\lambda_i-1}(C+A)^{\lambda_{23}-\lambda_i-1}\Big)=0\,,
\end{split}
\end{align}
where we have used the symmetry properties of $A,B,C$ in \eqref{ABC_sym}. In terms of $f^{\cdot\cdots}_-$ it becomes 
\begin{align}\label{generalOPE}
\begin{split}
&(\mathcal{F}^{1234}+\mathcal{F}^{2134}+\mathcal{F}^{1243}+\mathcal{F}^{2143})f^{1234}_-(A,B)\\
&+(\mathcal{F}^{1342}+\mathcal{F}^{3142}+\mathcal{F}^{1324}+\mathcal{F}^{3124})f^{1342}_-(B,C)\\
&+(\mathcal{F}^{1423}+\mathcal{F}^{4123}+\mathcal{F}^{1432}+\mathcal{F}^{4132})f^{1423}_-(C,A)=0\,.
\end{split}
\end{align}
We can determine the polynomial form of the functions $f^{\cdot\cdots}_-$ to be
\begin{subequations}\label{possible_functions}
\begin{align}
    f^{1234}_-(A,B)&=k^{1234}_-A^{\Lambda-2}-k^{1234}_+B^{\Lambda-2}\,,\\
    f^{1342}_-(B,C)&=k^{1342}_-B^{\Lambda-2}-k^{1342}_+C^{\Lambda-2}\,,\\
    f^{1423}_-(C,A)&=k^{1423}_-C^{\Lambda-2}-k^{1423}_+A^{\Lambda-2}\,.
\end{align}
\end{subequations}
By substituting these forms of the functions in \eqref{generalOPE}, we find
\begin{equation}
    A^{\Lambda-2}:(\mathcal{F}^{1234}+\mathcal{F}^{2134}+\mathcal{F}^{1243}+\mathcal{F}^{2143})k^{1234}_--(\mathcal{F}^{1423}+\mathcal{F}^{4123}+\mathcal{F}^{1432}+\mathcal{F}^{4132})k^{1423}_+=0\,,
\end{equation}
\begin{equation}
    B^{\Lambda-2}:(\mathcal{F}^{1342}+\mathcal{F}^{3142}+\mathcal{F}^{1324}+\mathcal{F}^{3124})k^{1342}_--(\mathcal{F}^{1234}+\mathcal{F}^{2134}+\mathcal{F}^{1243}+\mathcal{F}^{2143})k^{1234}_+=0\,,
\end{equation}
\begin{equation}
    C^{\Lambda-2}:(\mathcal{F}^{1423}+\mathcal{F}^{4123}+\mathcal{F}^{1432}+\mathcal{F}^{4132})k^{1423}_--(\mathcal{F}^{1342}+\mathcal{F}^{3142}+\mathcal{F}^{1324}+\mathcal{F}^{3124})k^{1342}_+=0\,.
\end{equation}
As usual, the solution for the couplings takes the form
\begin{equation}\label{usual_couplings}
    \mathcal{C}^{1234\lambda_i}=\frac{(k^{1234}_-+(-)^{\lambda_i+\lambda_{12}}k^{1234}_+)(\Lambda-2)!}{2^{\Lambda-2}(\lambda_{12}+\lambda_i-1)!(\lambda_{34}-\lambda_i-1)!}\quad
    \forall\;\lambda_i\,,\quad
    \text{same for $(1342)$ and $(1423)$}\,.
\end{equation}
The one presented here is a general solution, in the sense that it includes all possible non-zero terms and uses the most general form for the cubic vertices \eqref{cubic_vertex_general} in the OPE associativity constraint, with generic $\fA_{abc}$. 

As shown in the main text, the same solutions also satisfy the light-cone holomorphic constraint. The only difference is that this constraint leaves the products $C^{\lambda_1,\lambda_1,0}C^{0,\lambda_2,\lambda_2}$ unconstrained.

For instance, to recover the singlet case, we simply drop the $\mathcal{F}$ factors, which leads to the solution \eqref{commuting_case}. The sign differences arise from the different ordering of the external fields. Here, we have chosen the ordering that avoids additional signs.

Instead, if we want to recover the solutions for the colour case, for the colour-ordering $[1234]$, we need to consider the terms $\mathcal{F}^{1234}=\mathcal{F}^{4123}$ and will match with \eqref{colour_case}.

A complete analysis of the theories that satisfy the most general form of the constraint, along with a full classification, appears feasible but lies beyond the scope of this paper. As an application of the expression above, we solve the OPE associativity constraint in the mixed cases \eqref{mixed_case_1} and \eqref{mixed_case_2}. This allows us to identify certain admissible lower-spin chiral theories that would otherwise have been missed if one focused only on the singlet and colour cases.

\footnotesize
\providecommand{\href}[2]{#2}\begingroup\raggedright\endgroup

\end{document}